\documentclass[journal,10pt]{IEEEtran}
\usepackage[utf8]{inputenc}
\usepackage{cite}
\usepackage{graphicx,lipsum}
\usepackage{enumitem}
\usepackage[linesnumbered,ruled,vlined]{algorithm2e}
\usepackage{subcaption}
\usepackage{amsmath}
\usepackage{xcolor}
\usepackage{amssymb}
\usepackage[c3]{optidef}
\usepackage{algorithmicx}
\usepackage{algcompatible}
\usepackage{caption}
\usepackage{array}
\usepackage{pifont} 

\title{UAV-Assisted MEC Architecture for Collaborative Task Offloading in Urban IoT Environment\vspace*{-0.07in}}

\author{Subhrajit Barick, and Chetna Singhal,~\IEEEmembership{Senior Member,~IEEE}
\IEEEcompsocitemizethanks{\IEEEcompsocthanksitem
 S.~Barick is with the Department of Information and Communication Technology, Manipal Institute of Technology, Manipal Academy of Higher Education, Manipal, India and C.~Singhal is with INRIA, France.}\vspace*{-0.31in}}

\begin{document}
\maketitle

\begin{abstract}
Mobile edge computing (MEC) is a promising technology to meet the increasing demands and computing limitations of complex Internet of Things (IoT) devices. However, implementing MEC in urban environments can be challenging due to factors like high device density, complex infrastructure, and limited network coverage. Network congestion and connectivity issues can adversely affect user satisfaction. Hence, in this article, we use unmanned aerial vehicle (UAV)-assisted collaborative MEC architecture to facilitate task offloading of IoT devices
in urban environments. We utilize the combined capabilities of UAVs and ground edge servers (ESs) to maximize user satisfaction and thereby also maximize the service provider’s (SP) profit. We design IoT task-offloading as joint IoT-UAV-ES association and UAV-network topology optimization problem. Due to NP-hard nature, we break the problem into two subproblems: offload strategy optimization and UAV topology optimization. We develop a Three-sided Matching with Size and Cyclic preference (TMSC) based task offloading algorithm to find stable association between IoTs, UAVs, and ESs to achieve system objective. We also propose a K-means based iterative algorithm to decide the minimum number of UAVs and their positions to provide offloading services to maximum
IoTs in the system. Finally, we demonstrate the efficacy of the proposed task offloading scheme over benchmark schemes through
simulation-based evaluation. The proposed scheme outperforms by 19\%, 12\%, and 25\% on average in terms of percentage of
served IoTs, average user satisfaction, and SP profit, respectively, with 25\% lesser UAVs, making it an effective solution to support
IoT task requirements in urban environments using UAV-assisted MEC architecture. 
\end{abstract}

\begin{IEEEkeywords}
Mobile edge computing, Internet of Things, unmanned aerial vehicles, task offloading, user satisfaction, service provider profit.
\end{IEEEkeywords}
\vspace{-0.1in}
\section{Introduction}
The Internet of Things (IoT) technology supports numerous compelling applications with stringent requirements, such as real-time video analytics, augmented/virtual reality, video streaming, and automatic navigation \cite{Talebxr, video_streaming, automatic_navigation}. These applications heavily rely on quick data processing for intelligent decision-making. However, IoT devices have limited processing resources and battery life that make it challenging to execute computation-intensive tasks. Mobile Edge Computing (MEC) is a potential paradigm that addresses the computing limitations of IoT devices \cite{MEC_survey_comm_perspective,  MEC_Cooperative_taskOffloading_TNSM, MEC_cooperative_caching_TNSM}. By utilizing the computation and communication resources available at the network edge (e.g., cellular base stations, WiFi access points, and small-scale infrastructure), MEC enables IoT devices to execute computation-intensive applications by offloading them to the adjacent edge servers (ESs). This supports numerous advanced IoT applications by reducing service latency and improving service quality \cite{changUAV, three_sided_1, three_sided_2, neg_rcvdelay}. However, achieving dedicated wireless connections between IoTs and edge servers (ESs) may not always be feasible in practical scenarios. This challenge becomes more prominent in urban environments \cite{Urban_IoT_Environment}. The dense and varied cityscape, including tall buildings, creates physical barriers that hinder direct communication between IoT devices and edge resources. As a result, it becomes difficult to provide seamless connectivity and computing capabilities to all IoT devices. Deploying a large number of ESs across urban areas is a potential solution, but it requires careful planning, deployment, and management, which may increase the infrastructure implementation costs \cite{MEC_deployment_limitation}. To overcome this, Unmanned aerial vehicles (UAVs) are used to dynamically enhance the quality of service (QoS) of MEC in IoT networks.

UAVs support mobility, on-demand deployment \cite{barick2022multi,singhal_uav1}, and the ability to establish line-of-sight links \cite{singhal_uav3,singhal_uav6,ECMS_SUBHRAJIT}. 
Collectively UAVs can provide efficient task distribution, adapt to the dynamic user demands, and enhance service quality \cite{singhal_uav2,singhal_uav4,barick2022multi}. These attributes enable UAVs to support IoT devices in performing computation-intensive and delay-critical tasks. They can be utilized as computing platforms to assist IoT device tasks  \cite{UAV_assisted_MEC_SINGLE_STAGE_MAIN, UAV_assisted_MEC_MAIN_PAPER}. However, the UAVs have limited computational capacity, which may lead to significant delays while handling computational-intensive tasks \cite{challenges_UAV_assisted_MEC}. Furthermore, as the task load increases, it brings adverse effects on both system performance and UAV’s
battery lifetime \cite{UAV_ES_Payment1}. UAV-enabled MEC systems, with UAVs acting as relays, can be used to improve user experience (based on service latency \cite{three_sided_2}) in IoT networks by offloading the computational tasks to more resourceful ground ESs \cite{UAV_assisted_MEC_MAIN_PAPER, UAV_ASSISTED_TWO_STAGE}. Unlike fixed relays, UAVs can quickly move to optimal locations, providing improved coverage and line-of-sight communication, especially in areas where buildings or infrastructure obstruct signal transmission. This mobility allows UAVs to effectively handle dynamic offloading demands in urban environments, where IoT device density and distribution vary over time—challenges that would be difficult for ground-based infrastructure to address alone. 

\begin{figure}[h]
\centering
\frame{\includegraphics[width=3.3in]{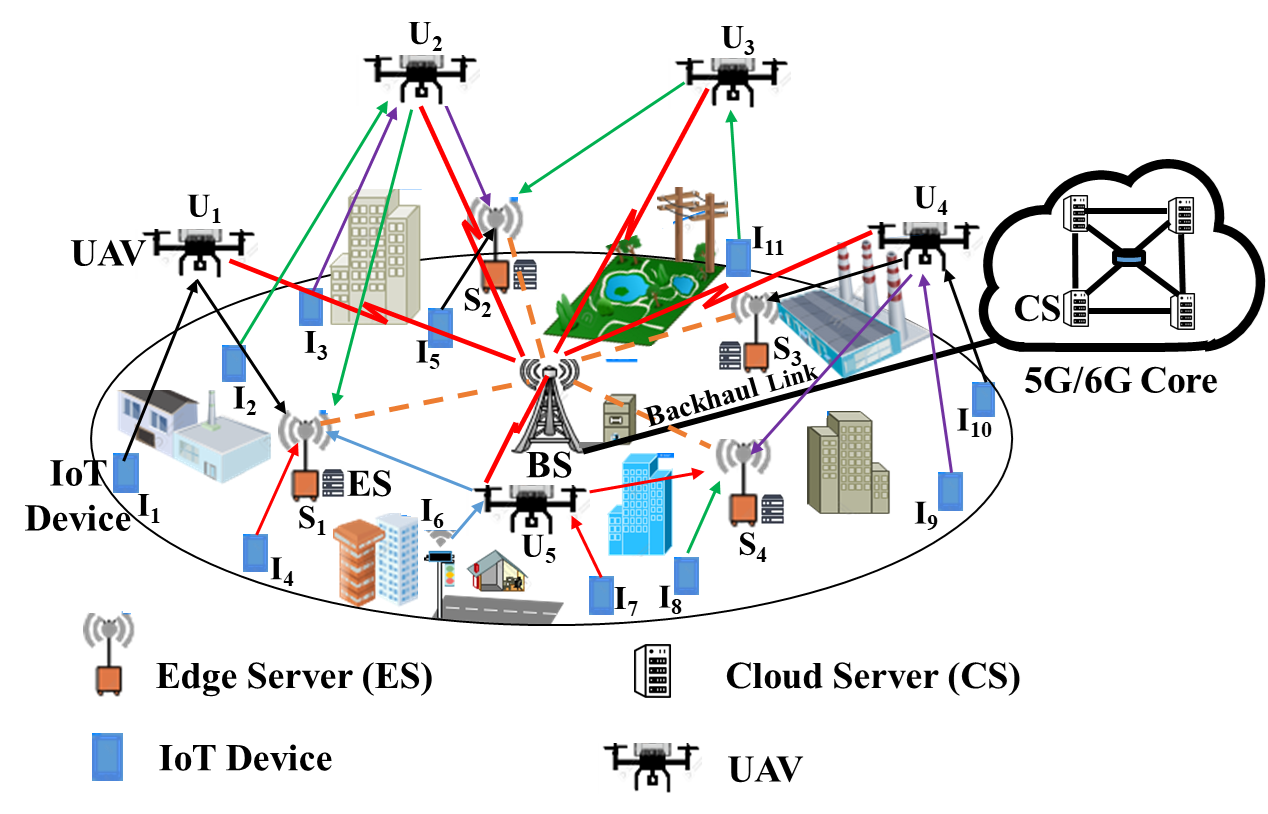}}
\caption{Multi-UAV assisted MEC system for IoT network}
\label{fig_sysmodel}
\vspace{-0.15in}
\end{figure}

However, in real-world networks, the edge resources are generally privately owned \cite{interplay_su_sp_ero} and the service provider (SP) has to lease these resources to support diverse IoT applications \cite{Economics_data_offloading}. Therefore, ensuring SP profit is equally important in UAV-assisted MEC networks along with user satisfaction \cite{SP_ES_Payment1}. Many recent studies in UAV-assisted MEC networks (\!\cite{UAV_assisted_MEC_MAIN_PAPER, challenges_UAV_assisted_MEC, UAV_ES_Payment1, UAV_ASSISTED_TWO_STAGE}) emphasize the optimization of user satisfaction while neglecting the importance of SP profit. Thus, in this work, we model a MEC architecture that prioritizes both user satisfaction and SP profit. We consider a multi-UAV assisted task offloading architecture with UAVs acting as relays working in collaboration with ground ESs to improve the performance of MEC systems in urban IoT networks. Fig. \ref{fig_sysmodel} shows a sample scenario of the proposed multi-UAV assisted MEC system for IoT networks.
To the best of our knowledge, UAV-assisted IoT task offloading in urban
environments involving collaboration between IoTs, UAVs, and ES owners, targeting to maximize the SP profit while meeting specific task requirements of IoTs has not been studied yet. Therefore, in this article, we propose an efficient task-offloading scheme to address this.

The key features of this article are summarized as follows:
\begin{enumerate}[leftmargin=*]
    \item A UAV-assisted MEC architecture to facilitate edge computing to IoT devices with limited connectivity. 
    
    \item IoT task offloading problem modeled as a joint IoT-UAV-ES association and topology optimization problem. 
    
    \item Optimization problem formulation to maximize SP profit while fulfilling task and service requirements.
    
    \item Three-sided Matching with Size and Cyclic preference (TMSC) and K-means based heuristic algorithm to find efficient UAV topology and stable IoT-UAV-ES association.

    \item Simulation-based performance evaluation with uniform and dataset based IoT distribution for varying numbers of IoTs and ESs, compares the proposed offloading with respect to state-of-the-art benchmark schemes in terms of percentage of served IoTs, number of UAVs deployed, average user satisfaction, and overall SP profit.
\end{enumerate}
The remaining article is structured as follows: Section II reviews of related works. Section III presents the system model and analyzes the SP profit using a revenue-cost model. We model the task offloading process as a joint optimization of UAV topology and IoT-UAV-ES association problem. Section IV discusses our algorithmic approach to solving the profit maximization problem. Section V presents the performance evaluation and effectiveness of the proposed offloading scheme while section VI concludes the article.\vspace{-0.1in}  
\begin{table}[t]
\centering

\tiny
\setlength{\tabcolsep}{2pt} 
\renewcommand{\arraystretch}{1.1} 
\begin{tabular}{|>{\raggedright\arraybackslash}p{0.8cm} | >{\raggedright\arraybackslash}p{1.8cm} | >{\centering\arraybackslash}p{1.0cm} | >{\centering\arraybackslash}p{0.8cm} | >{\centering\arraybackslash}p{0.6cm} | >{\centering\arraybackslash}p{0.6cm} | >{\centering\arraybackslash}p{0.8cm} | >{\centering\arraybackslash}p{0.5cm} | >{\centering\arraybackslash}p{0.35cm} |}
\hline
\textbf{References} & \textbf{Solution Approach} & \textbf{System-Cost, SP-Profit} & \textbf{Cyclic 3-sided Pref.} & \textbf{Adapt Real-Time} & \textbf{Multi-UAV Arch.} & \textbf{Urban network Scale} & \textbf{Latency} & \textbf{QoS} \\
\hline
\textbf{[7]}        & Reinforcement Learning                    & \ding{51} & \ding{55} & \ding{51} & \ding{55} & \ding{55} & \ding{51} & \ding{51} \\
\hline
\textbf{[6]}        & Deep Reinforcement Learning                    & \ding{51} & \ding{55} & \ding{55} & \ding{55} & \ding{55} & \ding{55} & \ding{51} \\
\hline
\textbf{[9]}       & Two-Sided Matching                        & \ding{55} & \ding{55} & \ding{55} & \ding{55} & \ding{55} & \ding{51} & \ding{51} \\
\hline
\textbf{[8]}        & Three-Sided Matching                      & \ding{51} & \ding{51} & \ding{51} & \ding{55} & \ding{55} & \ding{55} & \ding{51} \\
\hline
\textbf{[34]}  & Cyclic Stable Matching                    & \ding{51} & \ding{51} & \ding{55} & \ding{55} & \ding{51} & \ding{51} & \ding{51} \\
\hline
\textbf{[10]}       & Multi-Stage Offloading                    & \ding{55} & \ding{55} & \ding{55} & \ding{55} & \ding{51} & \ding{51} & \ding{51} \\
\hline
\textbf{[29], [32]} & Multi-Agent Deep Reinforcement Learning   & \ding{51} & \ding{55} & \ding{51} & \ding{51} & \ding{51} & \ding{51} & \ding{51} \\
\hline
\textbf{[33]}       & Fixed Offloading                          & \ding{51} & \ding{55} & \ding{55} & \ding{51} & \ding{51} & \ding{55} & \ding{51} \\
\hline
\textbf{[16], [27]} & Random Offloading                         & \ding{55} & \ding{55} & \ding{51} & \ding{51} & \ding{51} & \ding{51} & \ding{51} \\
\hline
\textbf{Our Approach}   & Three-Sided Matching w/ Cyclic Pref. & \ding{51} & \ding{51} & \ding{51} & \ding{51} & \ding{51} & \ding{51} & \ding{51} \\
\hline
\end{tabular}
\caption{Features Comparison between Proposed Strategy
and state-of-the-art schemes}\vspace{-0.3in}
\label{feature_compare}

\end{table}
\section{Related Works}
Task offloading helps to meet the computational requirements of resource-constrained heterogeneous IoT applications  \cite{MEC_Cooperative_taskOffloading_TNSM}. This involves parameters like offloading latency, energy consumption, and system cost \cite{MEC_cooperative_caching_TNSM, EC_min_delay3}. 
Existing solutions that minimize the task execution delay in a MEC system are based on deep reinforcement learning \cite{MEC_Cooperative_taskOffloading_TNSM} and optimization of service caching for multi-cell scenarios \cite{EC_min_delay3}. 
Collaboration between the edge nodes and BS reduces the end-to-end delay in cellular networks \cite{neg_rcvdelay}. A collaboration between edge node and cloud server can further improve the efficiency of task offloading in heterogeneous cellular network \cite{MEC_cooperative_caching_TNSM}. An incentive framework for optimizing the system cost of mobile data offloading in such a system has the ES operator different from the network SP, and the SP pays for the offloading traffic to the ES \cite{Economics_data_offloading}.  

However, these existing task-offloading strategies rely on consistent network connectivity between IoT devices and edge servers. In urban environments with high device density and complex infrastructure, network connectivity can be challenging \cite{Urban_IoT_Environment}. This can lead to delays in data transmission, task execution, and overall performance degradation \cite{MEC_limitations}. To overcome this, integration of UAVs
into MEC networks in urban environments can provide communication support and assistance in task offloading services \cite{UAV_assisted_MEC_MAIN_PAPER, sugg_1}. The optimization of UAV deployment, user association, and resource allocation can improve the average offloading latency of the system \cite{latency_single_UAV_1,singhal_uav5}. Additionally, the UAV trajectory and task allocations can be optimized to improve QoS of a MEC system \cite{energy_consumption_single_UAV_1}. Collaboration between UAV and edge cloud (EC) in the UAV-enabled MEC systems can enhance the system computation rate \cite{comp_rate_single_UAV_1} and  minimize service delay for IoT devices facing signal outage and shadowing  \cite{challenges_UAV_assisted_MEC}. 

A single UAV-assisted MEC solution only provides  connectivity over a small area due to its limited coverage, battery life, and payload capacity. Therefore a multi-UAV enabled MEC system is a more efficient and reliable alternative for task offloading in urban environments \cite{QoS_Improvement_multiUAV_MEC}. A multi-UAV architecture can support on-demand task execution of computation-intensive applications on IoT devices  \cite{multiUAV_taskOffloading_TNSM}. Additionally, it can meet the low latency and QoS requirement of IoT applications \cite{Talebxr, changUAV}. Collaboration between UAVs and BS helps in the efficient utilization of the available resources and improves the user experience in the MEC system \cite{UAV_BS_collaboration_multi_UAV}. Similarly, a collaboration between UAV and ESs can provide MEC services with reduced service latency to the IoT devices that are beyond the communication range of ECs by means of two-sided matching between the UAVs and ESs \cite{multi_UAV_UE_improvement_1} or a cooperative deep reinforcement learning approach \cite{multi_UAV_UE_improvement_2}. 

We note that the aforementioned studies have primarily focused on enhancing system performance from only the user’s perspective, neglecting the SP perspective. However, the SP being profit-driven, needs to maximize the revenue while ensuring acceptable QoS for users \cite{SP_revenue}. We remark that the SP’s perspective is equally important. Therefore, in this work, we propose a multi-UAV-assisted IoT edge-computing model considering both the user as well as the SP perspective. We have summarised 
the comparison of the proposed task offloading strategy with the other state-of-the-art schemes in terms of distinct features in Table \ref{feature_compare}. \vspace{-0.1in}

\section{System Model and Problem Formulation}
Fig. \ref{fig_sysmodel} shows the system model for the proposed multi-UAV assisted IoT edge computing architecture with an underlying cellular 5G network. The system comprises of one macro base station (MBS), $K$ UAVs, and $L$ ESs. These components collaboratively facilitate task offloading service to $M$ active IoT devices that may face difficulty in accessing ground ESs due to limited connectivity. Let $\mathcal{I} = \{ I_1, I_2, \dots, I_M \}$, $\mathcal{U} = \{ U_1, U_2, \dots, U_K\}$, and $\mathcal{S} = \{S_1, S_2, \dots, S_L\}$ denote the set of IoT devices, UAVs, and ESs in the system, respectively. Each active IoT device $I_m$ has a computation-intensive task represented by the tuple $T_m = \left(D_m, C_m, \tau_m\right)$, where $D_m$ (in bits) indicates the data size, $C_m$ (in CPU cycles/bit) is the computing resource requirement to process single-bit data, and $\tau_m$ (in seconds) denotes the maximum tolerable task completion delay. We denote the system-level IoT task
requirements as $\mathcal{T} = \{T_m, \forall m = 1, 2, \dots, M \}$. We consider that the tasks are offloaded to the ESs and not processed locally due to limited computing resources available at the IoT device. The UAVs are deployed as relays, which receive the task offloading requests from the IoTs and forward them to the appropriate ESs to meet IoTs' requirements. The ESs are deployed with the small cell 5G NR gNodeB that are privately owned (can support 5G
private networks). The network SP leases and pays (to the ES owner) for the computing resources of these ESs to perform task-offloading of the IoT devices \cite{SP_ES_Payment1, Economics_data_offloading}. Our primary objective is to maximize SP profit while ensuring seamless task offloading to a maximum number of IoTs, considering unique requirement of each IoT device.

We consider task offloading for a quasi-static scenario, where the operational duration is divided into numerous small intervals. Within each interval, we assume that the positions of IoTs and UAVs remain constant. We use a 3-D Cartesian coordinate system to represent the positions of IoTs, UAVs, and ESs within a small interval. The positions of IoT device $I_m$ and ES $S_l$ are denoted by $Q_{I_m} = (x_{I_m}, y_{I_m}, 0)$, and $Q_{S_l} = (x_{S_l}, y_{S_l}, 0)$, respectively, while the position of UAV $U_k$ is denoted as $Q_{U_k} = (x_{U_k}, y_{U_k}, H)$, where $H$ indicates the flying height of UAVs. Let $Q_I = \{Q_{I_m}, \forall m = 1, 2, \dots, M\}$, $Q_U = \{Q_{U_k}, \forall k = 1, 2, \dots, K\}$, and $Q_S = \{Q_{S_l}, \forall l = 1, 2, \dots, L\}$ be the set of positions of IoTs, UAVs, and ESs, respectively. The UAV altitude $H$ is predefined based on the communication coverage area on the ground while limiting the experienced path loss \cite{ECMS_SUBHRAJIT, barick2022multi}. The positions of UAVs $(x_{U_k}, y_{U_k})$ are determined using the proposed algorithm (discussed in section V). With this, the communication and task offloading models are explained as follows.

\begin{table}[!htp]
\label{T:equipos}
\begin{center}
\small
    \begin{tabular}{|p{0.5in}|p{2.6in}|}
    \hline
    \cline{1-2}
    \textbf{Symbol} & \textbf{Physical significance} \\
    \hline
    \!\!$M, K, L, Z$\!\!  &\!\!Number of IoT devices, UAVs, ESs, and PRBs, resp.\!\!\\
     \hline
    $\mathcal{I}$, $\mathcal{U}$, $\mathcal{S}$  & Set of IoT devices, UAVS, and ESs, resp.\\ \hline
    $T_m$ & Task generated at IoT device $I_m$   \\
    \hline
    $D_m$, $C_m$ & Data size, CPU cycle requirement of task $T_m$  \\
    \hline
    $\tau_m$ & Maximum tolerable delay of task $T_m$    \\
    \hline    $\mathbf{\delta}_{m,K,l}$ & Offloading strategy of IoT device $I_m$ \\
    \hline
    $\Delta$ & Set of Offloading strategies of all IoT devices  \\
    \hline
    $X_U$, $Y_U$ & Set of X- and Y- positions of UAVs\\
    \hline
    $q_l$ & Processing capacity of ES $S_l$\\
    \hline
    $\mathcal{F}$ & Set of processing capacities of ESs\\
    \hline
    $\mathcal{T}$ & Set of task requirements of all IoTs \\
    \hline
    $N_U$ &  IoT user capacity of UAV   \\
    \hline
    $N_S$ & Task processing capacity of ES  \\
    \hline
    \!\!$\Gamma_{m,k}$;  $\widetilde \Gamma_{k,l}$\!\!& SINR of $I_m$ at $U_k$; SINR of $U_k$ at $S_l$ \\
    \hline
    $P_m$ & Transmitting power of IoT device $I_m$ \\
    \hline
    $P_k$ & Transmitting power of UAV $U_k$ \\
    \hline
    $g_{m,k}$, $g_{k,l}$ & Channel gain between $I_m$ and $U_k$, $U_k$ and $S_l$ \\
    \hline
    $N_0$ & Noise power \\
    \hline
    $r_{m,k}$ & Data transmission rate between $I_m$ and $U_k$  \\
    \hline
    $r_{k,l}$ & Data transmission rate between $U_k$ and $S_l$ \\
    \hline 
    $B_1, B_2$ & Bandwidth assigned to IoTs and UAVs, respectively \\
    \hline   
    $t_{m,k,l}$ & Offloading delay associated with the strategy $\mathbf{\delta}_{m,k,l}$ \\
    \hline    $\mathcal{V}_{m,k,l}$ & User satisfaction associated with the strategy $\mathbf{\delta}_{m,k,l}$ \\
    \hline    $\mathcal{R}_{m,k,l}$ & Revenue earned for offloading strategy $\mathbf{\delta}_{m,k,l}$ \\
    \hline    $\mathcal{C}_{m,k,l}$ & Cost incurred by SP for offloading strategy $\mathbf{\delta}_{m,k,l}$ \\
    \hline    $\mathcal{P}_{m,k,l}$ & Profit earned for the offloading strategy $\mathbf{\delta}_{m,k,l}$ \\
    \hline\!\!$CP(\mathbf{\delta}_{m,k,l})$ & Cost-performance of offloading strategy $\mathbf{\delta}_{m,k,l}$    \\
    \hline
    \end{tabular}
\end{center}
\vspace{-0.1in}
\caption{Notations and significance of symbols}
\vspace{-0.2in}
\label{Notations}
\end{table}

\vspace{-0.1in}\subsection{Communication Model:} The communication between UAVs and ESs is facilitated by the MBS using $Z$ physical resource blocks (PRBs). These PRBs are equally shared among all UAVs. Additionally, each UAV uses $N_U$ PRBs (non-overlapping with MBS PRBs) to facilitate communication with the IoTs. We assume that each IoT device can be assigned a maximum of one PRB. Therefore, the maximum IoT user capacity of each UAV is considered as $N_U$. 

We use the probabilistic LoS model to represent data transmission between the ground components (IoTs or ESs) and UAVs \cite{np_hard}. We express the path loss between the ground component $i$ and UAV $j$ as 
\begin{equation}
    PL_{i,j} = \beta_0 d_{i,j}^2 \left[ P_{i,j}^{LoS}\mu_{LoS} + P_{i,j}^{NLoS}\mu_{NLoS}\right]
    \label{pathloss_all}
\end{equation}
where $\beta_0 = \left(\frac{4 \pi f}{c}\right)^2$, $f$ is the carrier frequency, $c$ is the speed of light,  $d_{i,j} = \sqrt{(x_i - x_j)^2 + (y_i - y_j)^2 + H^2}$ is the distance between the component $i$ and UAV $j$, $\mu_{LoS}$ and $\mu_{NLoS}$ are the additional pathloss components for LoS and NLoS links, respectively. $P_{i,j}^{LoS}$ and $P_{i,j}^{NLoS}$ are the probabilities corresponding to the LoS and NLoS links, respectively, and are given as\vspace{-0.1in}

\begin{equation}
    P_{i,j}^{LoS} = \frac{1}{1+a\exp{(-b(\theta_{i,j} - a))}}
\end{equation}
\begin{equation}
    P_{i,j}^{NLoS} = 1 - P_{i,j}^{LoS}
\end{equation}
where $a$ and $b$ are the environment constants, and $\theta_{i,j} = \sin^{-1}(\frac{H}{d_{i,j}})$ is the elevation angle of UAV $j$ with respect to component $i$. The data rate between the two communicating components $i$ and $j$ is governed by the link Signal to Interference Noise Ratio (SINR), which is given as:
\begin{equation}
    \Gamma_{i,j} = \frac{(P_i / PL_{i,j})}{\sum_{i' = 1, \, i' \neq i}^\eta  (P_{i'} / PL_{i',j}) + N_0},
    \label{SINR_all}
98\end{equation}
where $P_i$ is the transmitting power of $i$, $N_0$ is the noise power, and $\sum_{i' = 1, \, i' \neq i}^\eta  (P_{i'} / PL_{i',j})$ indicates the total interference power experienced at $j$ due to channel reuse. Since we focus on dense urban scenario, the PRBs may be reused across different clusters, 
and we carefully assign the PRBs to significantly reduce or even eliminate this co-channel interference under the given conditions. The task offloading rate is:
\begin{equation}
    r_{i,j} = B_{i,j} \log (1 + \Gamma_{i,j}),
\end{equation}
where $B_{i,j}$ is the communication bandwidth between ground component $i$ and UAV $j$. 

The SINR of IoT $I_m$ while offloading task to UAV $U_k$ is  $\Gamma_{m,k}$ with $\eta=M$ and the transmitting distance of $U_k$ from $I_m$ as $d_{m,k} = \sqrt{[(x_{U_k} - x_{I_m})^2 + (y_{U_k} - y_{I_m})^2 + H^2]}$. Here, $B_{i,j} = B_1$, corresponds to the bandwidth of PRB assigned to $I_m$ for offloading its task to $U_k$. Similarly, the SINR for UAV $U_k$ while relaying an offloaded task to ES $S_l$ is  $\Gamma_{k,l}$ with $\eta=K$ and $d_{k,l} = \sqrt{[(x_{U_k} - x_{S_l})^2 + (y_{U_k} - y_{S_l})^2 + H^2]}$, the transmitting distance of $U_k$ from $S_l$. Here, the UAV is assigned a bandwidth $B_{i,j} = B_2$ for relaying offloaded tasks to the ESs. The SINR corresponding to the IoT-UAV pair is measured at the UAV, while the SINR for the UAV-ES pair is measured at the ES \cite{UAV_ASSISTED_TWO_STAGE}. These SINR values are then shared with the MBS over separate control channels. We denote the set of SINR values for all possible IoT-UAV and UAV-ES pairs as $ \Gamma 
= \{ \Gamma_{m,k}, \forall m, k\}$ and $\widetilde \Gamma = \{ \Gamma_{k,l}, \forall k, l\}$ respectively.\vspace{-0.1in}

\subsection{Task Offloading Model:}
The IoT device initiates the task offloading process by transferring its computation-intensive and delay-sensitive task to the associated UAV. The UAV then forwards the task to a suitable ES for further processing. The ESs are heterogeneous entities that exhibit heterogeneity in terms of their computing capacities. We represent the computing capacities of the ESs as a set $\mathcal{F} = \{q_1, q_2, \dots, q_l, \dots, q_L \}$, where $q_l$ indicates the computing capacity (in CPU cycles/sec) of ES $S_l$. We assume that the computing resource of each ES is proportionally reserved for all the connected UAVs. The maximum number of tasks that each ES can handle is denoted as $N_S$ and can be determined by taking the ratio of the average ES CPU cycles to the average CPU cycle requirements of IoT tasks in the system. If the ES CPU cycle  capacity in the system is within the range [$q_{min}, q_{max}$], 
then $N_S$ is given as:
\begin{equation}
    N_S = \frac{q_{avg}}{D_{avg}\;C_{avg}}
\end{equation}
where $q_{avg}$, $D_{avg}$, and $C_{avg}$ are the average of ES CPU cycles, IoT task data size, and IoT task CPU cycle requirements, respectively, in the system.

We denote the offloading strategy for IoT device $I_m$ as $\delta_{m,k,l}$, where $\delta_{m,k,l}$ is defined as:
\vspace{-1mm}
\begin{equation}    \hspace{-0.09in}\mathbf{\delta}_{m,k,l} \!= \!\begin{cases}
       1, \;  \text{ if $T_m$ is offloaded to ES $S_l$ via UAV $U_k$}
        \\
        0, \;  \text{ otherwise}
        \end{cases}
\end{equation}
The overall offloading strategy for the system can be represented as $\Delta = \{\delta_{m,k,l} | \: \forall m = 1,\dots, M,\: k = 1,\dots, K,\: \text{and},\: l = 1,\dots, L,\}$. When selecting an offloading strategy, the SP aims to minimize the offloading cost while satisfying the strict delay requirements of IoT tasks. For a given strategy $\mathbf{\delta}_{m,k,l} = 1$, the offloading delay can be determined as \vspace{-0.05in} 
\begin{multline}
     \hspace{-0.3cm}t_{m,k,l} = t_{m,k,l}^{trans} + t_l^{proc}, \\
     \!=\! \left[\frac{D_{k,m}^{buffer}}{r_{m,k}} \!+\! \frac{D_m - D_{k,m}^{buffer}}{\min\left(r_{m,k}, r_{k,l}\right)} \!+\! \frac{D_{k,m}^{buffer}}{r_{k,l}}\right] + \frac{D_m\,C_m}{q_{l,m}},
     \label{delay_calculation}
\end{multline}
where $t_{m,k,l}^{trans}$ is the transmission delay when offloading the task $T_m$ from $I_m$ to $S_l$ via $U_k$ and $t_l^{proc.}$ is the processing delay when executing the task $T_m$ at $S_l$. Here, we are neglecting the result receiving delay from the ES to the IoTs, since the result’s data size is significantly smaller
compared to the input task size \cite{three_sided_2}. $D_{k,m}^{buffer}$ is the packet buffer size at UAV $U_k$ reserved for IoT $I_m$ (we consider that the total buffer space available at the UAV is reserved in equal proportion for all associated IoTs). $q_{l,m}$ denotes the computing resources at $S_l$ assigned for task $T_m$. The task offloading is successful, if $t_{m,k,l} < \tau_m$. We further analyze the SP's profit as follows.

\subsubsection{SP Profit}
The IoT users pay price for successful offloading of their computation-intensive tasks, which we consider as the revenue for the SP. We assume, similar to \cite{three_sided_1,three_sided_2}, that the price offered by an IoT user is directly proportional to the task size and inversely proportional to the delay deadline. Hence, the SP's revenue for an offloading decision $\mathbf{\delta}_{m,k,l} = 1$, is given as: 
\vspace{-1mm}
 \begin{equation}
    \mathcal{R}_{m,k,l} = \mathcal{P}_m = v \; \frac{D_m}{\tau_m},
    \label{price_offered_iot}
\end{equation}
\vspace{-1mm}
where $\mathcal{P}_m$ denotes the price paid by IoT $I_m$ and $v$ is the unit data rate price parameter (arbitarily and theoritically assumed to be dollar/Mbps). However, at the same time, the SP hires computing resources from the ES owners for task execution and pays for it. Similar to \eqref{price_offered_iot}, we assume that the cost paid by the SP depends
both on the task size as well as the processing capacity of the
hired ES \cite{three_sided_1}, \cite{three_sided_2}. Hence, the cost incurred by the SP for an offloading decision $\mathbf{\delta}_{m,k,l} = 1$ is given as:
\vspace{-1mm}
\begin{equation}
    \mathcal{C}_{m,k,l} = w \; \frac{D_m }{t_l^{proc}},
    \label{SP_cost_taskprocessing}
\end{equation}
\vspace{-1mm}
where $w$ [dollar/Mbps] is the price parameter \cite{three_sided_2}. The SP's profit for the offloading decision $\mathbf{\delta}_{m,k,l} = 1$, can be defined as:
\begin{equation}
    \mathcal{P}_{m,k,l} = \mathcal{R}_{m,k,l} -  \mathcal{C}_{m,k,l},
    \label{SP_profit_taskoffloading}
\end{equation}
Here, we consider $v \gg w$. Therefore, for any successful offloading (i.e, $\mathbf{\delta}_{m,k,l} = 1$), $\mathcal{R}_{m,k,l} > \mathcal{C}_{m,k,l}$. This implies, $\mathcal{P}_{m,k,l} > 0$, i.e., a positive SP profit is incurred for each successful task offloading. In addition to the task offloading cost, the SP has to incur a UAV operating cost which is proportional to the number of UAVs employed. Thus, given a specific network topology and a fixed offloading strategy, the overall SP profit can be expressed as: 
\begin{equation}
    \mathcal{P}_{SP} = \left[\sum_{m=1}^M\sum_{k=1}^K\sum_{l=1}^L \mathbf{\delta}_{m,k,l}\mathcal{P}_{m,k,l} - K\mathcal{C}^{UAV}\right]
    \label{SP_profit}
\end{equation}
where $\mathcal{C}^{UAV}$ represents the operational cost of a UAV. Eqs. \eqref{SP_cost_taskprocessing} and \eqref{SP_profit_taskoffloading} demonstrate how the offloading decision influences the SP profit, while \eqref{SP_profit} highlights the impact of the number of UAVs. According to \eqref{SP_cost_taskprocessing}, the SP incurs a higher cost in offloading a task (with a given requirement) to a more computationally powerful ES than a lesser one. Therefore, by strategically selecting the ES, the SP can complete task offloading at a minimal cost, maximizing its profit. Similarly, by optimizing the number of UAVs, the SP can reduce additional operating costs, thereby enhancing its overall profit.
\vspace{-0.3in}
\subsection{Problem Formulation}
The UAV-assisted IoT task offloading in urban IoT environments has two objectives, (1) Deliver offloading services with minimum UAV resources, and (2) Maximize the SP profit while meeting the IoT's requirements. Therefore, the task-offloading problem can be modeled as:\vspace{-0.05in}

{\small
\begin{maxi!}|s|
        {\mathbf{\Delta}, K, X_U, Y_U}{\mathcal{P}_{SP}}{}{P1:}       \addConstraint{\sum_{k=1}^K\sum_{l=1}^L\mathbf{\delta}_{m,k,l} \leq 1 \hspace{0.5cm} \forall m \in \{1,2,\dots, M\}}{\label{C1_P1}}        
        \addConstraint{\sum_{m =1}^M\sum_{l =1}^L\mathbf{\delta}_{m,k,l} \leq N_U \hspace{0.5cm} \forall k \in \{1,2,\dots,K\}}{\label{C2_P1}}
        \addConstraint{\sum_{m = 1}^M\sum_{k = 1}^K\mathbf{\delta}_{m,k,l} \leq N_S \hspace{0.5cm} \forall l \in \{1,2,\dots,L\}}{\label{C3_P1}}
        \addConstraint{\sum_{m = 1}^M\sum_{k = 1}^K\sum_{l=1}^L\mathbf{\delta}_{m,k,l} \geq \zeta M \hspace{0.5cm} }{\label{C4_P1}}
        \addConstraint{\Gamma_{m,k}, \Gamma_{k,l} \geq \Gamma_{th}}{\label{C5_P1}}
        \addConstraint{t_{m,k,l} \leq \tau_m \hspace{0.5cm}  \forall m \in \{1,2,\dots, M\}}{\label{C6_P1}}        \addConstraint{\mathbf{\delta}_{m,k,l} \in \{0,1\}}{\label{C7_P1}}
        \vspace{-0.05in}
\end{maxi!}
}
\vspace{-1mm}
where $X_U = \{x_{U_k}, \forall k = 1, 2, \dots, K\}$ and $Y_U = \{y_{U_k}, \forall k = 1, 2, \dots, K\}$ are the set of X-coordinates and Y-coordinates of the UAVs, respectively. The constraint (\ref{C1_P1}) specifies that each IoT can only be paired with only one UAV, and the corresponding task can only be offloaded to a single ES. Constraints (\ref{C2_P1}), and (\ref{C3_P1}) indicate the capacity constraints for the UAVs, and ESs, respectively. Constraint (\ref{C4_P1}) outlines the minimum service demand of the system. \eqref{C5_P1} and \eqref{C6_P1} highlight the SINR and delay criteria for selecting a suitable IoT-UAV-ES offload pair.
\vspace{-0.1in}
\section{Problem Analysis and Solution Design}
The optimization problem 
$P1$ is combinatorial and NP-hard due to binary constraints (13h) and the strong coupling between variables. To solve this effectively, we decompose $P1$ into two subproblems: offloading strategy optimization ($P2$) and UAV topology optimization ($P4$). 
Given an initial UAV topology, $P2$ seeks the best IoT-UAV-ES associations that satisfy SINR, delay, and capacity constraints. 
 Given the offloading strategy from $P2$, $P4$ adjusts the UAV network topology to maximize coverage while reducing UAV count. Next, we discuss these subproblems in further details. 
\vspace{-2mm}
\subsection{Offloading Strategy Optimization}
Given, a fixed UAV network topology, we can find the optimal IoT-UAV-ES associations by solving the following optimization problem. 
\vspace{-1mm}
\begin{maxi!}|s|
    {\mathbf{\Delta}}{\mathcal{P}_{SP}}{}{P2:}
    \addConstraint{\eqref{C1_P1}, \eqref{C2_P1}, \eqref{C3_P1}, \eqref{C5_P1}, \eqref{C6_P1}, \eqref{C7_P1}}{}
\end{maxi!}
Solving the above optimization problem is NP-hard \cite{np_hard}. Therefore, we adopt a preference-based stable matching model known as Three-sided Matching with Size and Cyclic Preference (TMSC) \cite{TMSC_2013}, to solve the problem. The TMSC model is used to handle scenarios having cyclic relationships among matching agents. In the context of our UAV-assisted MEC system, we observe a similar cyclic relationship among the communicating agents' preferences. The IoT devices offload their tasks to specific UAVs that provide better SINR. Therefore, IoT devices exhibit preferences only for UAVs. The UAVs forward the offloaded tasks to suitable ESs that maximize the SP's profit while satisfying the task requirements. So, the UAVs exhibit preferences over ESs only. Similarly, the ESs preferentially select IoTs that have less task computational complexity so that they can complete the task faster and accommodate more users, thus increasing the revenue. Hence, the preference lists of IoTs, UAVs, and ESs exclusively consist of UAVs, ESs, and IoT devices, respectively. This cyclic preference structure highlights the relevance of employing the TMSC model to solve our offloading strategy optimization problem $P2$ effectively. 

By using TMSC, we aim to find maximum number of matching triplets (Iot-UAV-ES pairs) that maximize the system objective while satisfying the system constraints. The set of matching triplets is denoted as $\mathcal{M}_t = \{(I_m, U_k, S_l) |\: \mathbf{\delta}_{m,k,l} = 1 \; \forall I_m \in \mathcal{I}, \forall U_k \in \mathcal{U}, \forall S_l \in \mathcal{S}\}$. Each successful task offload corresponds to a unique triplet ($I_m, U_k, S_l$) in $\mathcal{M}_t$ that yields a positive profit for the SP, i.e. $\mathcal{P}_{m,k,l} > 0$. Therefore, for a given network topology (i.e., with fixed operation cost), we can maximize the SP profit by maximizing the number of matching triplets. Hence, we can reformulate the problem $P2$ as follows:
\vspace{-1mm}
\begin{maxi!}|s|
    {\mathbf{\Delta}}{ |\mathcal{M}_t|}{}{P3:}       
    \addConstraint{\mathcal{N}(\mathcal{M}_t, U_k) \leq N_U \hspace{0.4cm}\forall k \in \{1,2,\dots, K\}}{\label{C1_P3}}
    \addConstraint{\mathcal{N}(\mathcal{M}_t, S_l) \leq N_S \hspace{0.5cm}\forall l \in \{1,2,\dots, L\}}{\label{C3_P3}}
    \addConstraint{\mathcal{N}(\mathcal{M}_t, I_m) \leq 1 \hspace{0.7cm}\forall m \in \{1,2,\dots, M\}}{\label{C4_P3}}
    \addConstraint{t_{m,k,l} \leq \tau_m \hspace{1.4cm}  \forall m \in \{1,2,\dots, M\}}{\label{C5_P3}}
    \addConstraint{\Gamma_{m,k}, \Gamma_{k,l} \geq \Gamma_{th}}{\label{C6_P3}}            
    \vspace{-0.05in}
\end{maxi!}
The objective is to maximize the number of matched triplets in the matching set $\mathcal{M}_t$. $\mathcal{N}(\mathcal{M}_t, x)$ represents the number of times agent $x$ appears in the matching set $\mathcal{M}_t$. Constraints \eqref{C1_P3} and \eqref{C3_P3} represent the limitation on the maximum number of IoTs supported by the UAV and ES, respectively. The constraint \eqref{C4_P3} ensures that each IoT device should form only a single triplet, i.e, the IoT device $I_m$ can only be associated with one UAV and its task can be offloaded to a single ES. The matching triplets must comply with the maximum delay and minimum SINR constraints, as indicated by condition \eqref{C5_P3} and \eqref{C6_P3}, respectively. 

Finding a stable matching set for a given instance of TMSC is an NP-hard problem \cite{TMSC_2013}. Therefore, we reduce the computational complexity by imposing certain restrictions into the TMSC model and devise a Restricted-TMSC (R-TMSC) based algorithm to solve the task offloading problem. The preference lists of all agents are derived from the master preference lists available at the MBS. The master list is a collection of agents with strict order in the form of a single or a group of lists. The master preference list of IoTs is a set of $M$ lists, where each list corresponds to a distinct IoT device and is a collection of all UAVs, arranged in the descending order of their SINR value, $\Gamma_{m,k}$. Similarly, the master preference list of UAVs is a single list having all the ESs arranged in the ascending order of their task computing capacity, $q_l$. The master preference list of ESs contains all the IoTs arranged in the ascending order of their task complexity, $D_m C_m$. The preference lists of individual IoT, UAV, and ES are prepared by including all or just part of the master list agents.  

The preference lists of IoTs are generated by including the UAVs from the master list that have SINR greater than the minimum SINR required for efficient data transmission (i.e., $\Gamma_{th}$). The preference list is defined as\vspace{-0.05in} 
\begin{multline}
    \hspace{-0.4cm}PL_{I_m}\!=\!\{ (u_1, u_2, \dots, u_K) | u_k = U_j, u_{k'}\!=\!U_{j'}, k\!<\! k^{'}, \forall \, k, k^{'} \\\in [1, K],\Gamma_{m,j} \geq \Gamma_{m,j^{'}}\geq \Gamma_{th},  \forall \, U_j, U_{j'} \in \mathcal{U}, j \neq j^{'}\} 
    \label{pref_IoT}
\end{multline}
$\forall m \in \{1,2,\dots, M\}$. The set of preference lists of all IoTs is denoted as $PL_I = \{ PL_{I_m}, m = 1,2,\dots, M \}$. Similarly, the preference lists of UAVs can be derived using Eq. (\ref{PL_UAV}) based on the master list that ranks all the ESs according to their processing capacity.\vspace{-0.05in} 
\begin{multline}
    \hspace{-0.4cm}PL_{U_k} = \{ (s_1, s_2, \dots, s_L) | s_l = S_v, s_{l'} = S_{v'}, l < l^{'}, q_v \leq q_{v'}, \\ \hspace{-0.1cm}\forall \, l, l^{'} \in [1,L], \forall \, S_v, S_{v'} \in \mathcal{S}, \,v \neq v^{'} \,\text{and} \, q_{v,k}, q_{v^{'},k} > 0 \}\! 
    \label{PL_UAV}
\end{multline}
$\forall k \in \{1,2,\dots, K\}$. Let $PL_U = \{ PL_{U_k}, k = 1,2,\dots, K \}$ be the set of preference lists of all UAVs in the system. The ESs' preference lists is defined as:\vspace{-0.05in}
\begin{multline}
    \hspace{-0.3cm}PL_{S_l} = \{ (i_1, i_2, \dots, i_M) | i_m = I_r, i_{m'} = I_{r'}, m < m',D_{r}C_{r} \\ \leq D_{r'}C_{r'}, \forall m, m' \in [1,M], \forall I_r, I_{r'} \in \mathcal{I}, \text{and} \,r \neq r^{'}\}\!
    \label{PL_ES}
\end{multline}
$\forall l \in \{1,2,\dots, L\}$. $PL_S = \{ PL_{S_l}, l = 1,2,\dots, L \}$ is the set of preference lists of all ESs in the system. With the knowledge of preference lists, we propose a R-TMSC motivated algorithm to find stable matching triplets that satisfy both task requirements and SP's objectives.

\subsubsection{R-TMSC based Matching Algorithm}
Algorithm 1 is the matching algorithm that solves the problem $P3$. Given a matching $\mathcal{M}_t$ and any triplet ($I_m, U_k, S_l$) $\in \mathcal{M}_t$, we denote $\mathcal{M}_t$($I_m$) = $U_k$, $\mathcal{M}_t$($U_k$) = $S_l$, and $\mathcal{M}_t$($S_l$) = $I_m$. Let 
\begin{equation}    \hspace{-0.2cm}\scalebox{0.87}{$\mathcal{Z}_{I_m} \!= \!\{U_k | U_k \succ  \mathcal{M}_t(I_m), U_k \in PL_{I_m}, \text{and} \: \mathcal{N}(\mathcal{M}_t, U_k) < N_U\}$},\hspace{-0.25cm} 
    \label{UAV_SET_ZIm}
\end{equation}
be the set of UAVs that are preferred by IoT $I_m$ over its current match $\mathcal{M}_t(I_m)$ and have available resources (PRBs) to provide offloading services. When $\mathcal{M}_t(I_m) = \phi$, we set $\mathcal{Z}_{I_m} = PL_{I_m}$. Similarly, 
\begin{equation}
    \scalebox{0.95}{$\mathcal{Z}_{U_k} = \{S_l | S_l \succ  \mathcal{M}_t(U_k), \text{and}\: S_l \in PL_{U_k}\}$},
\end{equation}
denotes the set of ESs that the UAV $U_k$ prefers to its current match $\mathcal{M}_t(U_k)$. For $\mathcal{M}_t(U_k) = \phi$, we set $\mathcal{Z}_{U_k} = PL_{U_k}$.  
\begin{multline}            \hspace{-0.3cm}\scalebox{0.95}{$\hat{\mathcal{Z}}_{I_m}\! =\! \{S_l | I_m \in PL_{S_l}, \mathcal{N}(\mathcal{M}_t, S_l) < N_S, \text{and}\: t_{m,k,l} < \tau_m  \}$},\hspace{-0.3cm}
\end{multline}
represents the set of ESs, which have the capacity to accept the task request from $I_m$ while satisfying the delay requirement of the task $T_m$. Then, 
\begin{equation}
    \scalebox{0.95}{$\mathcal{Z}_{I_m}^* = \{U_k : \mathcal{Z}_{U_k} \cap \hat{\mathcal{Z}}_{I_m} \neq \phi, \text{and}\: \mathcal{N}(\mathcal{M}_t, U_k) \leq N_U\}$},
    \label{UAV_SET_ZIm_star}
\end{equation}
indicates the set of UAVs, for which there exists an ES in their preference list that can accept the task request from $I_m$. These sets are created (and maintained) at the BS and are updated whenever the matching set gets updated. The matching triplets of IoTs, UAVs, and ESs are derived using Algorithm \ref{alg:resorce_allocation}. 

Algorithm \ref{alg:resorce_allocation} takes input as the IoT, UAV, and ESs' position information ( $Q_I, Q_S, X_U, Y_U, H$), IoTs' task requirement information (\textbf{$\mathcal{T}$}), and ESs' computation resource ($\mathcal{F}$) information. It returns the offloading strategy ($\Delta$) as output. The algorithm first generates preference lists for each system agent (i.e., $PL_I, PL_U, PL_S$). These lists are then provided as arguments to the matching function TMSC\_MATCH, which produces a stable matching set $\mathcal{M}_t$ using an iterative approach. The function TMSC\_MATCH starts with an empty matching set $\mathcal{M}_t$ and progressively updates it during each iteration. In each iteration, the \emph{for} loop finds better agents (UAVs and ESs) for the IoTs and updates the matching set $\mathcal{M}_t$ with the better triplets. For any IoT $I_m$, $U^* \neq \phi$ indicates that a better triple exists. $U^*$ and $S^*$ are the set of UAVs and ESs, respectively that $I_m$ prefers over its current match. $Head(U^*, I_m)$ and $Head(S^*, U_k)$ select the highest priority elements in the list $U^*$ and $S^*$, with respect to $I_m$ and $U_k$ respectively. Finally, the better triple is added to the matching set $\mathcal{M}_t$ by removing the existing one. This process repeats until we obtain a stable matching set. This stable matching set $\mathcal{M}_t$ is an input for the OFFLOAD\_STRATEGY function that produces the offload strategy $\Delta$ based on the elements of $\mathcal{M}_t$.

\begin{algorithm}[!htb]
\small
\caption{\hspace{-0.02in}R-TMSC based offloading strategy\!\!}
\label{alg:resorce_allocation}
\begin{algorithmic}
\State \textbf{Input:}$Q_I, Q_S, X_U, Y_U, H, \mathcal{T}, \mathcal{F}$\\
 \SetKwFunction{FMain}{Main}
  \SetKwFunction{FSum}{\!\!}
  \SetKwFunction{FSub}{\!\!}
\begin{enumerate}
    \item [1)] Prepare preference list sets $PL_{I}, PL_{U}, PL_{S}$ 
\end{enumerate}
        
\noindent\SetKwProg{Fn}{Function I: TMSC\_MATCH}{:}{}
  \Fn{\FSum{$PL_I, PL_U, PL_S$}}{
   \begin{enumerate}     
      \item [1)] {\bf Initialize} $\mathcal{M}_t = \phi$, and\: $flag = 1$;
 \end{enumerate}
    \begin{enumerate}     
      \item [2)] {\bf Find matching triplets}
 \end{enumerate}
        ${}$\hspace{1.8em}\While {$flag == 1$}{
            $flag = 0$;\\
            \For{$m = 1$ to $M$}{
                 \hspace{-2.0em}$U^*$ = $\mathcal{Z}_{I_m} \cap \mathcal{Z}_{I_m}^*$;\\
                \If{$U^* \neq \phi$}{
                      \hspace{-2.0em}$U_k$ = $Head(U^*)$;\\
                     $S^*$ = $\mathcal{Z}_{U_k} \cap \hat{\mathcal{Z}}_{I_m}$;\\
                     $S_l$ = $Head(S^*)$;\\
                    \If{$\mathcal{N}(\mathcal{M}_t, I_m)$ == 1}{
                         \hspace{-1.0em}\scalebox{0.9}{$\mathcal{M}_t$\,=\, $\mathcal{M}_t \setminus \{I_m, \mathcal{M}_t(I_m), \mathcal{M}_t(\mathcal{M}_t(I_m))\}$;}
                   }
                   \hspace{-2.0em}$\mathcal{M}_t$ = $\mathcal{M}_t \cup \{I_m, U_k, S_l\}$\\
                        $flag$ = 1;
                    }
                   
                }
            }
        \KwRet{$\mathcal{M}_t$}
        }
        
\noindent\SetKwProg{Fn}{Function II: OFFLOAD\_STRATEGY}{:}{}
    \Fn{\FSum{$\mathcal{M}_t$}}{
    \begin{enumerate}     
     \item [1)] {\bf Find offloading strategy}
    \end{enumerate}
    ${}$\hspace{2.0em}\For{$m = 1$ to $M$}{\\
    \hspace{1.0em}\For{$k = 1$ to $K$}{\\
    \For{$l = 1$ to $L$}{
        \hspace{-0.5cm}\eIf{$\{I_m, U_k, S_l\} \in \mathcal{M}_t$}{\hspace{-0.7cm}    $\mathbf{\delta}_{m,k,l} = 1$}{\hspace{-0.5cm}$\mathbf{\delta}_{m,k,l} = 0$
        } 

    }
    }
    }
    \KwRet{$\mathbf{\Delta}$}
    }
\end{algorithmic}
\vspace{-0.07in}
\end{algorithm}

\subsection{UAV topology Optimization}
The UAV topology optimization problem can be given as:\vspace{-0.05in}
\begin{maxi!}|s|
        {K, X_U, Y_U}{\mathcal{P}_{SP}}{}{P4:}           
        \addConstraint{\eqref{C2_P1}, \eqref{C4_P1}, \eqref{C7_P1}}{\label{C1_P4}}
        \vspace{-0.05in}
\end{maxi!}
Note that finding an optimal solution for the problem $P4$ is difficult due to coupling between the variables. The number of UAVs explicitly depends upon the UAV locations, and conversely, the positioning of the UAVs is influenced by the number of UAVs. Therefore, to solve the problem, we propose a K-means-based iterative algorithm. We start with the current UAV topology and iteratively update the number of UAVs till the constraints are met. Once we find the number of UAVs, the positions of UAVs are determined using the K-means method.    

K-means ensures that the UAV is positioned at the central location within the cluster, allowing it to effectively serve all users within its vicinity with minimum delay. Hence, for a cluster $CL_k$, the respective UAV location ($x_{U_k},y_{U_k}$) can be determined by solving $P5$.\vspace{-0.05in}
\begin{mini}|s|
        {x_{U_k},y_{U_k}}{\sum_{I_m \in \mathcal{I}_{CL_k}}\left( x_{U_k} - x_{I_m}\right)^2 + \left( y_{U_k} - y_{I_m}\right)^2,}{}{P5:}
        \label{UAV_location}
\end{mini}
where $\mathcal{I}_{CL_k}$ denotes the set of IoTs included in the cluster $CL_k$. The above problem is convex and the optimal solution can be obtained by equating
the first-order derivative to zero  \cite{barick2022multi}. So, the UAV location becomes:\vspace{-0.05in}
\begin{equation}
    \hspace{-0.3CM}\small x_{U_k} = \frac{1}{N_{CL_k}}\sum_{I_m \in \mathcal{I}_{CL_k}}x_{I_m}, \; and \;  y_{U_k} = \frac{1}{N_{CL_k}}\sum_{I_m \in \mathcal{I}_{CL_k}}y_{I_m} 
    \label{optimal_position_SU}
\end{equation}
where $N_{CL_k}$ is the number of IoTs in the cluster $CL_k$. After finding the horizontal locations, the UAVs are deployed at predefined height $H$ at these locations. The details of topology updation is given in the Algorithm \ref{alg:UAV_Topology}. The algorithm finds a feasible UAV topology solution that ensures $\zeta$ percent IoTs are served with their offloading requirements. In order to optimize the number of UAVs, we iteratively update the number of UAVs based on the number of served IoTs. When the number of served IoTs exceeds the threshold limit, we try to optimize the number of UAVs by eliminating the one whose elimination does not affect the system performance (i.e., the served IoT percentage does not go below the threshold limit). For this, we first identify the UAV serving the least number of IoTs, and then, eliminate the same after checking the constraint. This step is repeated till the number of UAVs converges. Similarly, when the number of served IoTs is less than the threshold limit, we increase the UAV count by a minimum number that is sufficient to meet $\zeta$ percent service criteria. $ max(1,\lfloor \frac{M - |S_{serve}|}{N_U}\rfloor)$ gives the additional number of required to serve the non-served users. After finding the number of UAVs, we decide their positions using K-means method. 

\begin{algorithm}[!htb]
\small
\caption{\hspace{-0.02in}Iterative Algorithm for UAV Topology\!\!}
\label{alg:UAV_Topology}
\begin{algorithmic}
    \State \textbf{Input:}$\mathbf{\Delta}, K, \zeta$\\
    \SetKwFunction{FMain}{Main}
    \SetKwFunction{FSum}{\!\!}
    \SetKwFunction{FSub}{\!\!}
    \noindent\SetKwProg{Fn}{Function I: UAV\_Topology}{:}{}
    \Fn{\FSum{{$\mathbf{\Delta}, K, \zeta$}}}{
    \begin{enumerate}     
        \item [1)] {\bf Initialize} $flag = 1$;
        \item [2)] {\bf Update} $\mathcal{S}_{serve} = \{I_m | \mathbf{\delta}_{m,k,l} = 1, \forall k, l\}$;
        \item [3)] {\bf Update number of UAVs and their positions}
    \end{enumerate} 

    \While {$flag == 1$}{
        $flag = 0$;\\
    \hspace{0.5cm}\eIf{$|S_{serve}| > \zeta M$}{\For{$k = 1$ to $K$}{
        \State\parbox[t]{\dimexpr\linewidth-\algorithmicindent}{\scalebox{0.93}{\hspace{0.4cm}$S_{serve}^k = \{I_m | I_m \in S_{serve}, \mathbf{\delta}_{m,k,l} = 1, \forall m, l\}$}};
        }
        $\hat{k} = \arg \min (|S_{serve}^k|)$;\\ \hspace{0.6cm}\eIf{$(|S_{serve}| - |S_{serve}^{\hat{k}}|) > \zeta M$}{\hspace{-0.1cm}$K = K - 1;$\\ \hspace{0.5cm}$flag = 1;$\\ \hspace{0.5cm}$S_{serve} = S_{serve} \setminus \ S_{serve}^{\hat{k}};$}{$K = K$;}
        }{
        $K = K + max(1,\lfloor \frac{M - |S_{serve}|}{N_U}\rfloor)$}
    }
    \For{$k = 1$ to $K$}{\\
         \hspace{1.8em}Compute $x_{U_K} \text{and}\; y_{U_K}$ using \eqref{optimal_position_SU}; 
    }\\
    \KwRet{$K, X_U, Y_U$}
    }
    
\end{algorithmic}
\end{algorithm}

Algorithm \ref{alg:iterative_Taskoffload} shows the joint optimization of UAV topology and offloading strategy for efficient task offloading. We start with an initial UAV topology. The initial topology should be selected such that the algorithm converges with minimum iterations. So, we assume that each UAV serves at its maximum capacity, i.e, $N_U$, and estimate the initial number of UAVs as
\begin{equation}
    K_{initial} = \biggl\lfloor\frac{M}{N_U}\biggr\rfloor
    \label{numUAV_initial}
\end{equation}
where $\lfloor.\rfloor$ is the floor operator. With the initial topology, we iteratively update the offloading strategy and the topology, till the profit conveges.

\begin{algorithm}[!htb]
\small
\caption{\hspace{-0.02in}Joint Optimization of Topology and Offloading Strategy for Profit Maximization\!\!}
\label{alg:iterative_Taskoffload}
\begin{algorithmic}
    \State \textbf{Input:}$Q_I, Q_S, N_U, H, f, c, \mu_{LoS}, \mu_{NLoS}, \zeta, a, b, $\\
    \begin{enumerate}
        \item [1:] Determine initial no. of UAVs ($K_{initial})$ using \eqref{numUAV_initial};
        \item [2:] Find initial UAV positions using \eqref{optimal_position_SU};
        \item [3:] {\bf Initialize} $K = K_{initial}, iter = 0, flag = 1$;
        \item [4:] {\bf Set} $\mathcal{P}_{SP}(iter) = 0;$
        \item [5:] {\bf Update} UAV topology and offloading strategy iteratively;
    \end{enumerate}
    ${}$\hspace{-0.3cm}\While {$flag == 1$}{
    $flag = 0$;\\
    \hspace{0.65cm}$iter = iter + 1$;\\
     \hspace{0.65cm}Update $\Delta$ using Algorithm \ref{alg:resorce_allocation};\\
     \hspace{0.65cm}Find $\mathcal{P}_{SP}(iter)$ using \eqref{SP_profit};\\
     \hspace{0.65cm}\eIf{$\mathcal{P}_{SP}(iter) - \mathcal{P}_{SP}(iter-1) < \epsilon$}{\scalebox{0.995}{$\Delta^* = \Delta, K^* = K, X_U^* = X_U, Y_U^* = Y_U$}}{Update $K, X_U$ and $Y_U$ using Algorithm \ref{alg:UAV_Topology};\\
     \hspace{0.65cm}$flag = 1$;}

    }
    \State \textbf{Output: $K^*, X_U^*, Y_U^*, \text{and}\;\Delta^*$}
\end{algorithmic}
\end{algorithm}
Algorithm \ref{alg:resorce_allocation} involves sequential matching of the preference list elements. Therefore, in the worst case where the elements are matched one by one, the algorithm has a computational complexity of $n_1 \times \mathcal{O}(MKL)$, where $n_1$ is the number of iterations that the algorithm \ref{alg:resorce_allocation} takes to converge. Similarly, algorithm \ref{alg:UAV_Topology} has a complexity order of $n_2 \times \mathcal{O}(K)$ with $n_2$ being the number of iterations required for convergence of algorithm \ref{alg:UAV_Topology}. So, if algorithm \ref{alg:iterative_Taskoffload} takes $n$ iteration to converge, then the overall computational complexity of the proposed algorithm can be expressed as $n\times 
[\mathcal{O}(n_{1}MKL) + \mathcal{O}(n_{2}K)]$.

\vspace{-0.06in}\section{Performance Analysis}
A simulation-based comprehensive performance analysis of our proposed task offloading scheme is carried out using MATLAB R2022b. We evaluate the effectiveness of our proposed strategy in comparison to two benchmarks: Fixed  \cite{multi_UAV_UE_improvement_2} and Random offloading \cite{UAV_assisted_MEC_MAIN_PAPER}. In fixed offloading, the IoT-UAV-ES associations are fixed, wherein the IoT device first offloads its task to the corresponding cluster UAV, which then forwards it to the nearest ES for execution. On the other hand, 
the random offloading allows task offloading without any predefined association plan. Together, these bench-
marks provide a balanced perspective on performance across both controlled and dynamic urban MEC settings.  We keep resource allocation consistent across all schemes for a fair comparison, and analyze the impact of different IoT-UAV-ES association strategies on system performance.

\vspace{-0.1in}\subsection{Simulation Framework}
We create a simulation environment for an area of 1km$^2$. The BS is placed at the center of the region and the IoT devices are distributed uniformly randomly across the region \cite{barick2022multi}. The ESs are located at predetermined fixed positions to facilitate the MEC services for IoT devices \cite{UAV_assisted_MEC_MAIN_PAPER}. To assist MEC service provision, the UAVs are deployed at a predetermined altitude \cite{ECMS_SUBHRAJIT}. The location of UAVs are determined using the K-means algorithm \cite{barick2022multi}. The parameter settings for the simulation are summarized in Table \ref{simulationparameter} \cite{UAV_assisted_MEC_MAIN_PAPER, UAV_assisted_MEC_SINGLE_STAGE_MAIN, three_sided_1, three_sided_2, ECMS_SUBHRAJIT}, \cite{multi_UAV_UE_improvement_2}. The task computing capacity of the ESs ($N_S$) are selected within the range [10, 50] tasks/s, and the price parameters are set as, $v=0.1$ and $w=0.01$ dollars/Mbps. 

\begin{table}[!htp]
\label{T:equipos}
\begin{center}
\small
\begin{tabular}{ |p{2.1in}|p{0.9in}|} 
 \hline
\cline{1-2}
\textbf{Parameters} & \textbf{Values} \\
\hline
UAV operating height (H) & 100 m \\ 
\hline
Power gain at reference distance ($\beta_o$) & $10^{-3}$\\
 \hline
Threshold SINR ($\Gamma_{th}$) & 3dB\\
\hline
Noise power ($N_0$) & -170 dBm  \\ 
\hline
IoT transmit power ($P_m$) & [100, 500] mW  \\ 
\hline
UAV transmit power ($P_k$) & 1 W\\
\hline
Data size of task ($D_m$) & [2, 5] MB \\
\hline
CPU cycle requirement of task ($C_m$) &\!\!\![100,200] cycles/bit\!\!\\ 
\hline
Maximum task completion delay ($\tau_m$)  & [2, 5] s\\ 
\hline
Task arrival rate & 30 tasks/min \\
\hline
Computing resource of ES ($q_l$) & [8, 10] GHz \\
\hline
Bandwidths of PRB ($B_1$) & 180 KHz  \\ 
\hline
Number of PRBs ($Z$) & 100 \\
\hline
UAV to ES transmission bandwidth ($B_2$)& 5 MHz  \\ 
\hline
Maximum IoT user capacity of UAV ($N_U$)\!\!& [10, 20] \\
 \hline
\end{tabular}
\end{center}
\vspace{-0.15in}
\caption{Simulation parameters for performance evaluation }
\vspace{-0.1in}
\label{simulationparameter}
\end{table}
With these simulation settings, we first evaluate the convergence of the proposed task offloading algorithm with $N_U\!=\!20$ and $N_S\!=\!40$. Then, we perform Monte-Carlo simulation with a 95 \% confidence interval for the configurations listed in table \ref{simulation data set}. We evaluate the system performance in terms of percentage of served IoTs, SP profit, and average user satisfaction. Since the proposed offloading aims to meet strict delay requirements of IoT tasks, we consider offloading latency for measuring user satisfaction\cite{three_sided_2}. For offloading strategy $\delta_{m,k,l} = 1$, we model the user satisfaction as a logarithmic function \cite{uav_offload_1}, given as:\vspace{-0.05in} 
\begin{equation}
    \mathcal{V}_{m,k,l} =  \!\begin{cases}
       \log \left(1 + \frac{\tau_m}{t_{m,k,l}}\right), \;  \text{ if $ 0 < t_{m,k,l} < \tau_m$}
        \\
        0, \;  \text{ otherwise}
        \end{cases}.
    \label{user_satisfaction}
\end{equation}   \vspace{-0.03in} 

\begin{table}[!htp]
\label{T:equipos}
\begin{center}
\begin{tabular}{|c|c|c|c}
\cline{1-3}
\textbf{Configurations} & \textbf{$|\mathcal{I}|$}          & \textbf{$|\mathcal{S}|$}     
\\ \cline{1-3}
Set 1                  & 100, 150, . . , 300 & 8                 & \\ \cline{1-3}
Set 2                  &200 & 2, 4, . . , 10  \\ \cline{1-3}

\end{tabular}
\end{center}
\vspace{-0.15in}
\caption{Experiment configurations. Set 1 and 2 has increasing number of IoTs and ESs, respectively.}
\vspace{-0.15in}
\label{simulation data set}
\end{table}

\vspace{-0.15in}\subsection{Performance Analysis}
\subsubsection{Algorithm convergence and UAV requirement analysis} 
\begin{figure}[!ht]
    \centering
        \centering
        \begin{subfigure}{0.48\columnwidth} 
            \includegraphics[width=\textwidth]{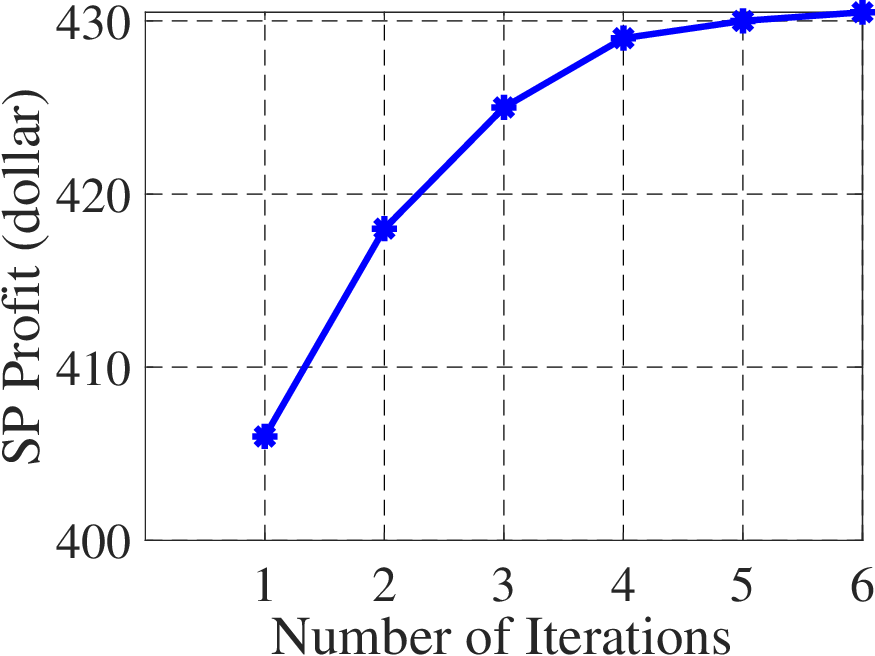}
            \vspace{-0.2in}
            \caption{Convergence performance of proposed algorithm with $M\!=\!200$,\; $L\!=\!8$,\; $N_U\!=\!20$,\; $N_S\!=\!40$.}
            \vspace{-0.1in}
            \label{fig:first}
        \end{subfigure}
        \hfill
        \begin{subfigure}{0.48\columnwidth} 
            \includegraphics[width=\textwidth]{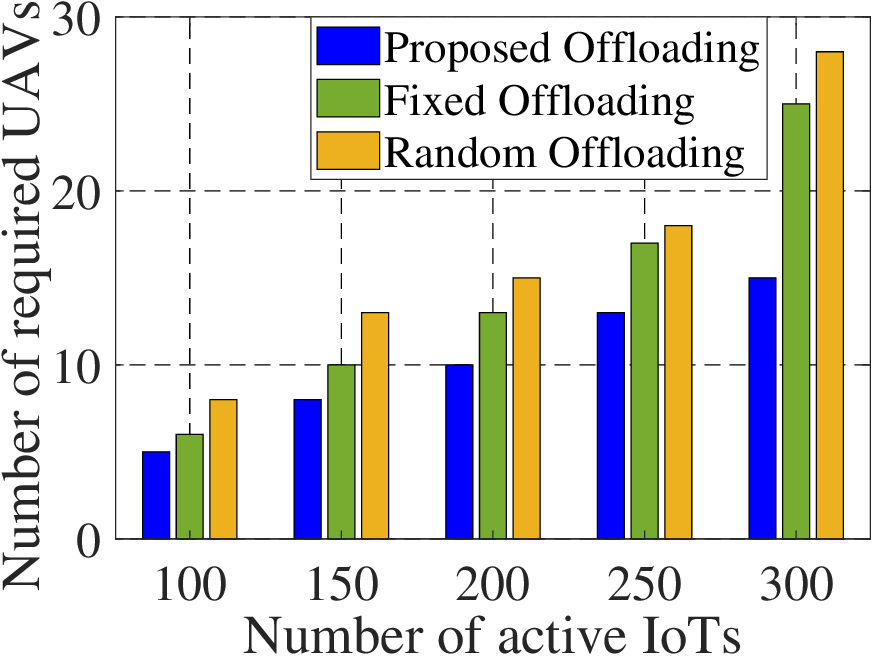}
            \vspace{-0.2in}
            \caption{Number of required UAVs with increasing number of active IoTs for $\zeta\!=\!90\%$}
            \vspace{-0.1in}
            \label{fig:second}
        \end{subfigure}
        \vspace{0.05in}
\caption{Algorithm convergence and UAV requirement analysis}
\vspace*{-0.12in}
\label{convergence and UAV requirement}
\end{figure}

Fig. \ref{convergence and UAV requirement}(a) shows the convergence pattern of the proposed iterative algorithm with $M\!=\!200$, $L\!=\!8$, $N_U\!=\!20$ and $N_S\!=\!40$. The SP profit increases with the number of iterations, and stabilizes just after four iterations. Each iteration executes in 18 ms and the algorithm converges within 60 ms on a Lenovo ThinkPad E14 with an Intel i5 processor. Overall, deploying the proposed solution in urban environments incurs a small ($<60\,ms$) execution-time overhead.

The minimum number of required UAVs with the proposed and the benchmark schemes is shown in Fig. \ref{convergence and UAV requirement}(b). We set the minimum served IoT percentage threshold as $\zeta = 90\%$ and evaluate the number of UAVs required under configuration set 1, with parameters $N_U\!=\!20$ and $N_S\!=\!40$. The number of required UAVs is greater for a higher number of IoTs. However, our proposed scheme effectively serves the required number of IoTs with fewer UAVs as compared to the benchmark schemes, thereby improving the SP profit.

\begin{figure}[!ht]
    \centering
        \centering
        \begin{subfigure}{0.48\columnwidth} 
            \includegraphics[width=\textwidth]{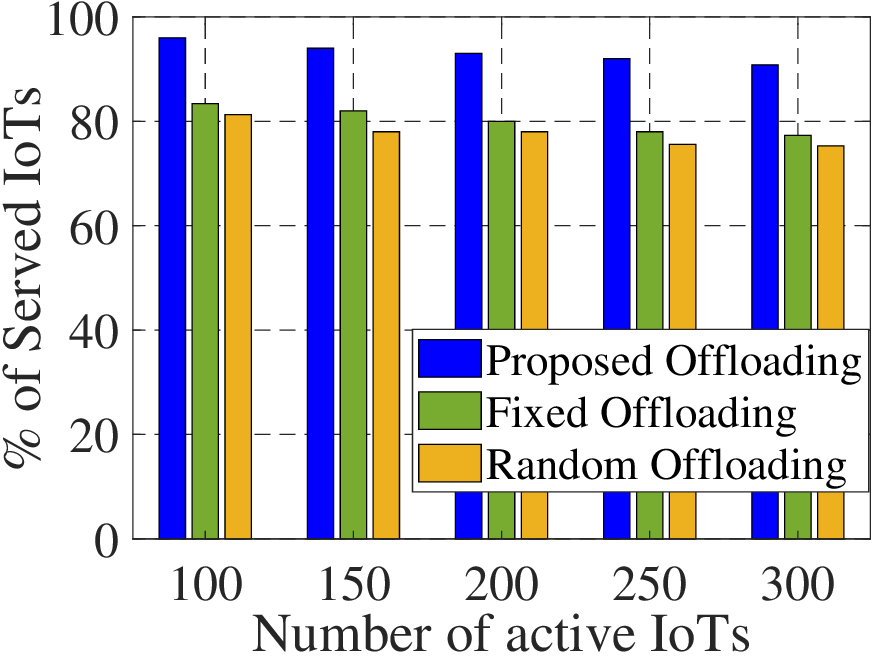}
            \vspace{-0.2in}
            \caption{Served IoT percentage}
            \vspace{-0.25in}
            \label{fig:first}
        \end{subfigure}
        \hfill
        \begin{subfigure}{0.48\columnwidth} 
            \includegraphics[width=\textwidth]{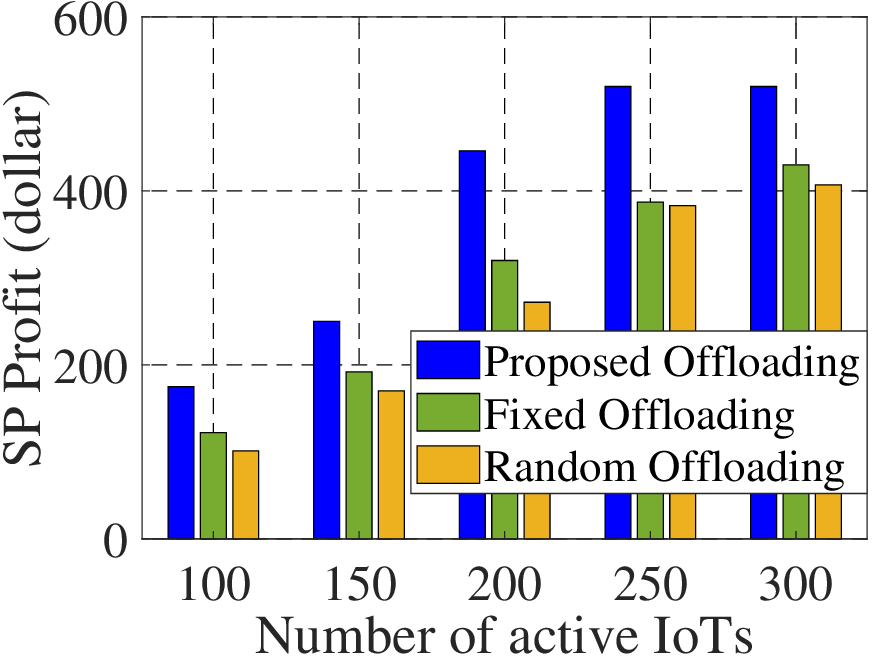}
            \vspace{-0.2in}
            \caption{Analysis of SP's profit}
            \vspace{-0.25in}
            \label{fig:third}
        \end{subfigure}
        \vspace{0.2in}
\caption{Performance with increase in number of active IoTs}
\vspace{-0.12in}
\label{analysis_of_IoT}
\end{figure}\vspace{-0.1in}

\vspace{0.3cm}\subsubsection{Effect of increasing number of active IoTs} The system performance for an increase in the number of active IoTs using configuration set 1 is shown in Fig. \ref{analysis_of_IoT}. 
Fig. \ref{analysis_of_IoT}(a) shows the percentage of served IoTs, that decreases with an increase in the number of active IoTs. This is due to an increased competition between the devices to obtain their preferred triplets with a higher number of active IoTs in the system. It may leave some of the devices out of pairing due to the capacity constraints of UAVs and ESs, leading to a decrease in the served percentage. Fig. \ref{analysis_of_IoT}(b) shows the effect on SP profit. The SP profit increases as the number of active IoTs increases. However, the rate of growth gradually slows down. This is because as the number of active IoTs increases, there is a saturation point after a certain threshold number
of IoTs (currently at 250) where the system reaches
its maximum capacity. Beyond this point, further increase in the number of active IoTs has minimal impact on the SP profit. 


We note 
that our proposed offloading strategy 
achieves an average served percentage of 93.4 \%, much higher than 80.1 \% for Fixed and 77.6 \% for Random offloading schemes. 
Also, the SP profit with the proposed offloading is 21\% and 28\% higher than the fixed and random strategies, respectively. This performance improvement with our scheme is attributed to the preference-based task offloading and the provision for less preferred pairs when more preferred choices are unavailable.

\begin{figure}[!ht]
    \centering
        \centering
        \begin{subfigure}{0.48\columnwidth} 
            \includegraphics[width=\textwidth]{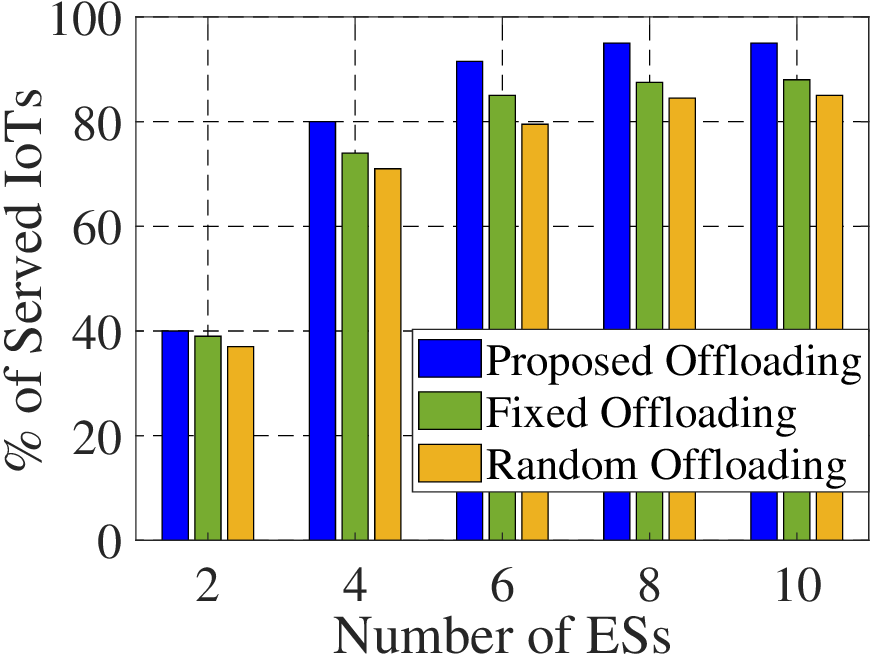}
            \vspace{-0.2in}
            \caption{Served IoT percentage}
            \vspace{-0.25in}
            \label{fig:first}
        \end{subfigure}
        \hfill
        \begin{subfigure}{0.48\columnwidth} 
            \includegraphics[width=\textwidth]{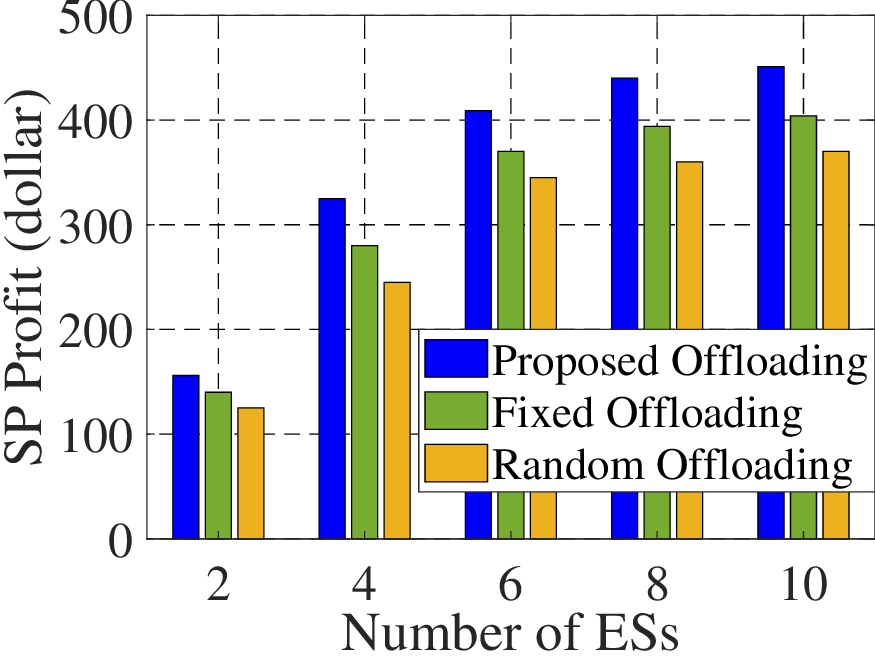}
            \vspace{-0.2in}
            \caption{Analysis of SP's profit}
            \vspace{-0.25in}
            \label{fig:third}
        \end{subfigure}
        \vspace{0.17in}
\caption{Performance with increase in number of ESs}
\vspace{-0.12in}
\label{analysis_of_ES}
\end{figure}

\subsubsection{Effect of increasing number of ESs}
Fig. \ref{analysis_of_ES} shows system performance with an increase in the number of ESs, based on the configuration set 2. We fix number of UAVs as $K = K_{initial} = 10$, evaluated using \eqref{numUAV_initial} with $M = $200, and $N_U = $20. 
Fig. \ref{analysis_of_ES}(a) shows the percentage of served IoTs that initially increases due to the formation of more matching triplets. However, due to fixed UAV capacity, the system becomes saturated after a certain number of ESs (8 in this case). Therefore, adding more ESs beyond this point does not lead to any significant improvement in system performance, resulting in a constant served IoT percentage. Fig. \ref{analysis_of_ES}(b) provides insights into the SP's profit. The graph shows the SP profit increases with the number of ESs. This is because as the number of ESs increases, the chance of UAVs being matched with their preferred ESs also increases, which reduces the task execution cost and thereby leads to an improvement in the SP's profit. However, the rate of growth in the performance metric becomes slower beyond 8 ESs, indicating that the system approaches a saturation point.  
Overall, the performance improvement with increasing number of ESs is constrained by the capacity of UAVs. This can be considered to determine an appropriate number of ESs (in this case it is 8) to achieve a balance between the benefits of increasing the number of ESs and the constraints of the UAV's capacity.
\begin{figure}[!ht]
    \centering
        \begin{subfigure}{0.48\linewidth}
            \includegraphics[width=\textwidth]{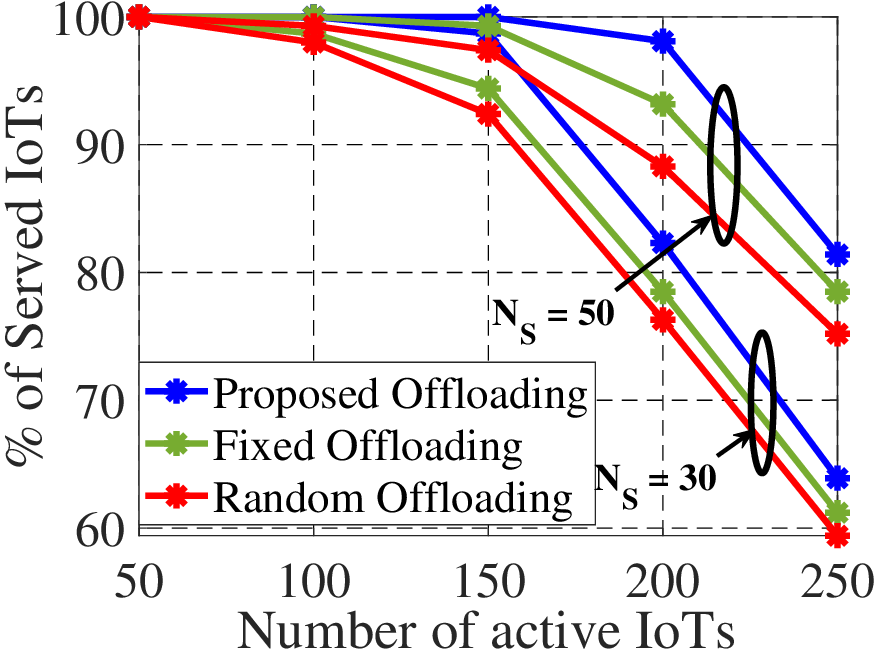}
            \caption{Effect of ES's computing capacity ($N_S$)}
            \label{fig:first}
            \vspace{-0.12in}
        \end{subfigure}
        \hfill
        \begin{subfigure}{0.49\linewidth}
            \includegraphics[width=\textwidth]{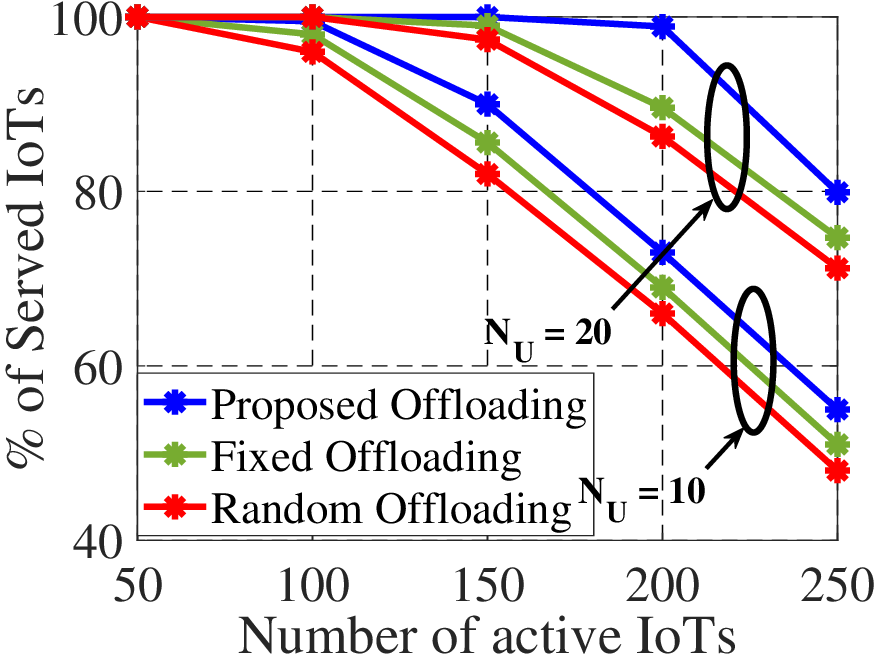}
            \caption{Effect of UAV's serving capacity ($N_U$)}
            \label{fig:second}
            \vspace{-0.12in}
        \end{subfigure}      \vspace{0.15cm}
\caption{Server and UAV capacity impact on performance.}\vspace{-0.15in}
\label{analysis_of_capacity}
\end{figure}

\subsubsection{Effect of the server and UAV capacity}
Fig. \ref{analysis_of_capacity} shows the system performance with different ES's computing and UAV's serving capacities for the configuration set 1. 
In Fig. \ref{analysis_of_capacity}(a), the percentage of served IoTs is shown for two different task computing capacities of ES: $N_S$ = 30, and $N_S$ = 50, while maintaining the IoT user capacity of UAVs at $N_U$ = 15. Similarly, Fig. \ref{analysis_of_capacity}(b) shows the performance for two different UAVS's serving capacity values: $N_U$ = 10, and $N_U$ = 20, while keeping $N_S$ = 40. The served IoT percentage increases as we increase the ES and UAV capacities that results in more matching triplets being formed, allowing more IoTs to successfully offload their tasks. 
Hence, for a given UAV-assisted MEC system, we can employ higher-capacity UAVs and ESs to improve the overall system performance. 

\begin{figure}[!ht]
    \centering
        \begin{subfigure}{0.48\linewidth}
            \includegraphics[width=\textwidth]{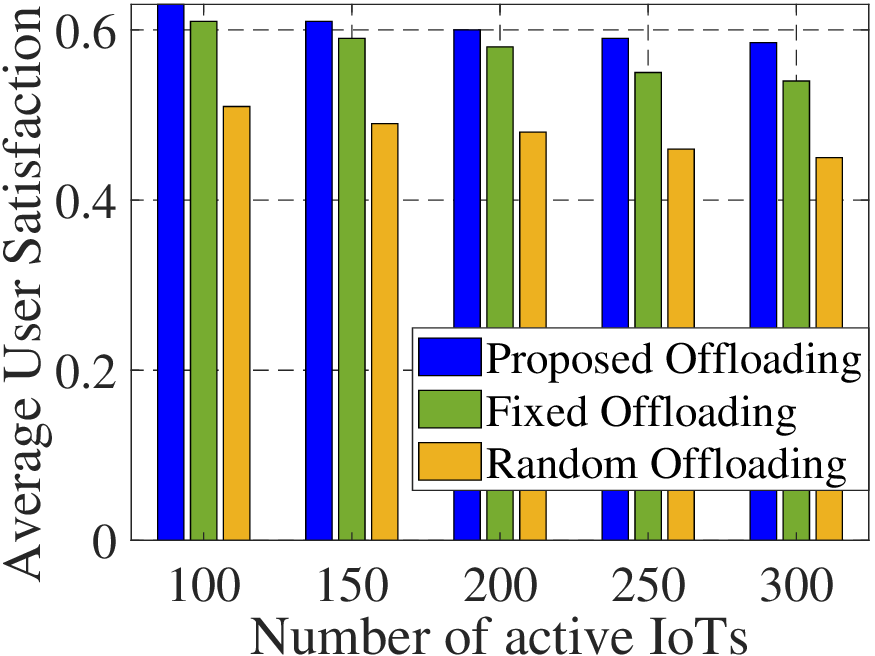}
            \caption{Configuration set 1}
            \label{fig:first}
        \end{subfigure}
        \hfill
        \begin{subfigure}{0.49\linewidth}
            \includegraphics[width=\textwidth]{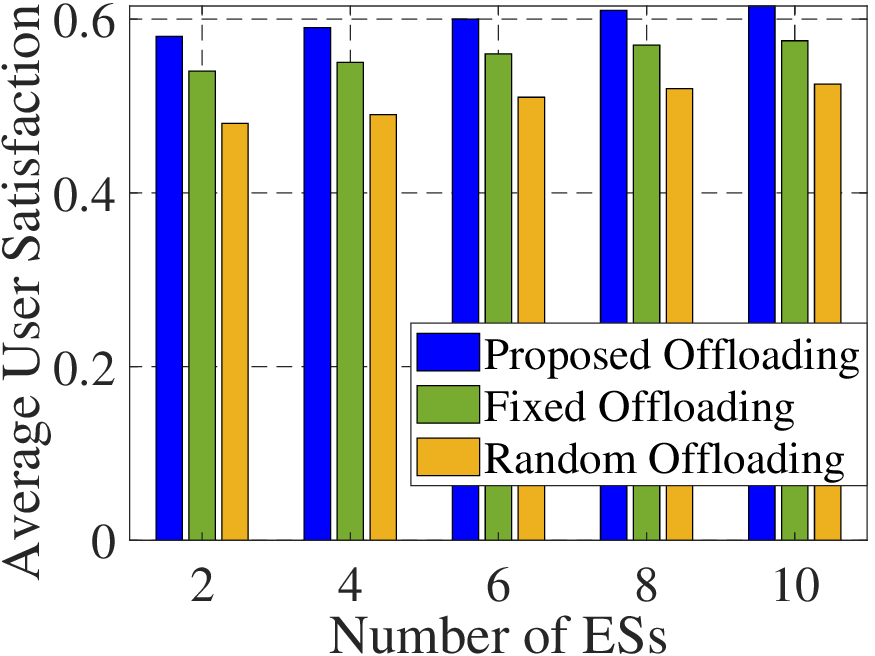}
            \caption{Configuration set 2}
            \label{fig:second}
        \end{subfigure}         

\caption{User satisfaction under different configuration setup.}
\label{analysis_of_user_satisfaction}
\vspace{-0.12in}
\end{figure}

\subsubsection{User satisfaction Performance}
Fig. \ref{analysis_of_user_satisfaction} shows the average user satisfaction performance for the configuration sets 1 and 2. 
We note from Fig. \ref{analysis_of_user_satisfaction}(a) that as the number of IoTs increases, the average user satisfaction decreases due to an increased competition among devices to obtain their preferred pairs (UAVs and ESs). However, the reduction rate becomes less pronounced beyond 250 IoTs, which can be attributed to the saturation state of the system due to the capacity constraints at the UAV and the ES level. Figs. \ref{analysis_of_user_satisfaction}(b) shows the variation of average user satisfaction with the increasing number of ESs. As the number of ESs increases, the average user satisfaction increases, which is due to the increased system capacity that allows the UAVs to offload the tasks to their preferred ESs. Furthermore, the increase in the number of ESs decreases the effective distance between the UAV and its preferred ES. This reduction in distance subsequently decreases the task transmission delay, which ultimately improves the average user satisfaction.  We notice from Figs. \ref{analysis_of_user_satisfaction}(a) and \ref{analysis_of_user_satisfaction}(b) that our proposed offloading approach delivers better user satisfaction than the benchmark
schemes. 

\begin{figure}[!ht]
    \centering
        \centering
        \begin{subfigure}{0.48\columnwidth} 
            \includegraphics[width=\textwidth]{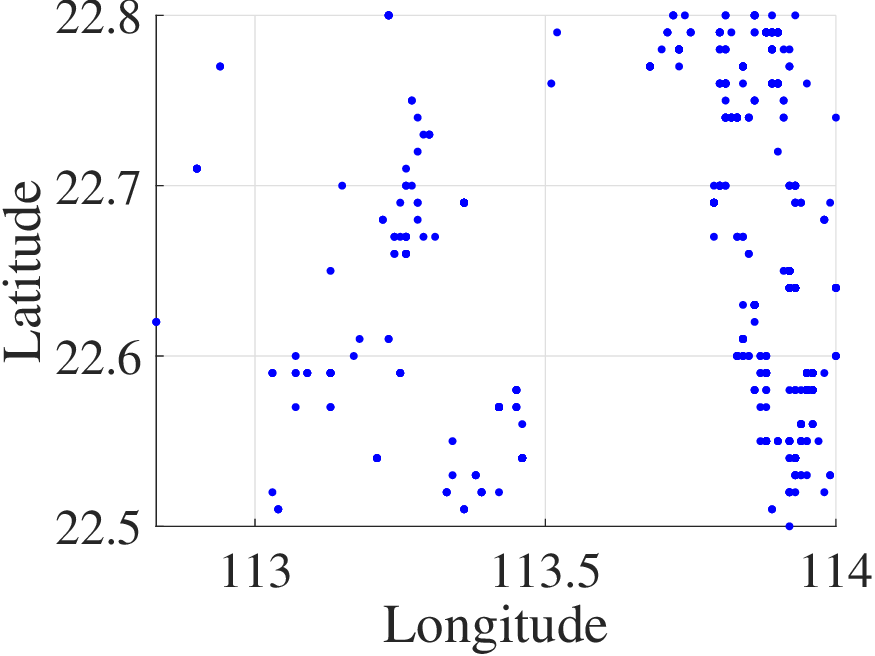}
            \vspace{-0.2in}
            \caption{User distribution in urban scenario.}
            \vspace{-0.2in}
            \label{fig:first}
        \end{subfigure}
        \hfill
        \begin{subfigure}{0.48\columnwidth} 
            \includegraphics[width=\textwidth]{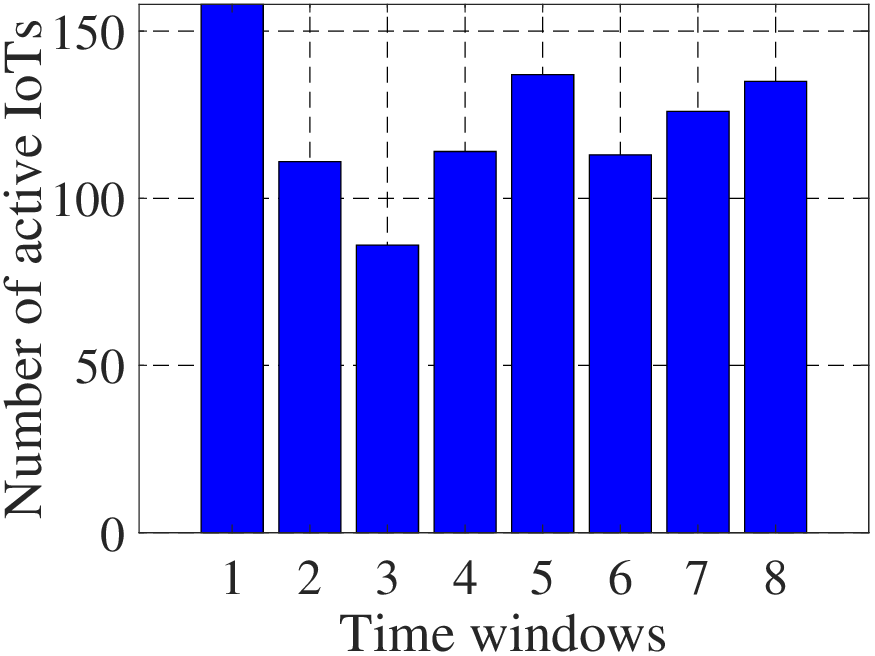}
            \vspace{-0.2in}
            \caption{Histogram of number of active IoTs across time windows}
            \vspace{-0.2in}
            \label{fig:third}
        \end{subfigure}
        \vspace{0.15in}
\caption{Simulation scenario based on real-world data}
\vspace*{-0.12in}
\label{Deployment_scenario}
\end{figure}\vspace{-0.05in}
\subsubsection{System performance based on real-world data}
In the above simulation, we have assumed a uniform distribution of IoT devices across the desired area. However, real-world user distribution may not always follow this uniform pattern. Therefore, to ensure the robustness and applicability of our algorithm, we aim to validate our system performance using real-world data. For this, we consider the \texttt{China Mobile User Gemographics} data set that provides information about mobile user activities, including timestamps and geographical coordinates \cite{china_mobile_gemographics}. We select a densely populated region bounded by latitude 22.5 to 23.3, and longitude 112.5 to 114.8. Fig. \ref{Deployment_scenario}(a) shows the mobile user distribution in this region between  10 AM to 12 PM. Based on this user distribution, we deploy 8 ESs at strategic locations to support task offloading services. The locations of these ESs are (22.775,113.8), (22.725,113.85), (22.675,113.85), (22.6, 113.85), (22.55, 113.85), (22.55, 113.4), (22.6, 113.2), and (22.7, 113.25). We divide the entire duration into 15-minute time windows and analyze the system performance across each window. The histogram in Fig. \ref{Deployment_scenario}(b) illustrates the variation in number of active IoTs over each window. 

\begin{figure*}[!ht]
    \centering
        \begin{subfigure}{0.3\textwidth}
            \includegraphics[width=\textwidth]{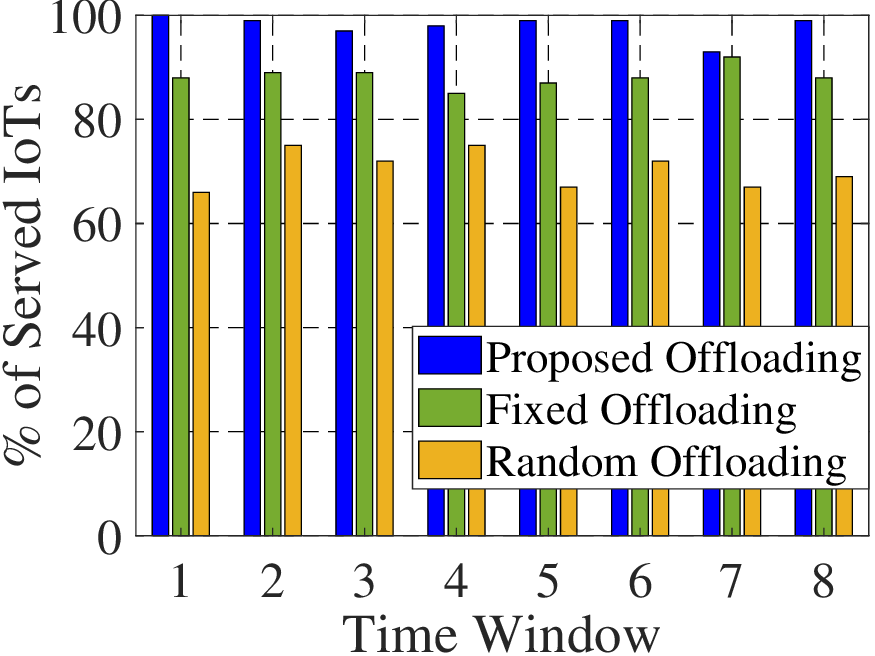}
            \vspace{-0.2in}
            \caption{Analysis of served IoT percentage}
            \vspace{-0.1in}
            \label{fig:first}
        \end{subfigure}
        \hfill
        \begin{subfigure}{0.3\textwidth}
            \includegraphics[width=\textwidth]{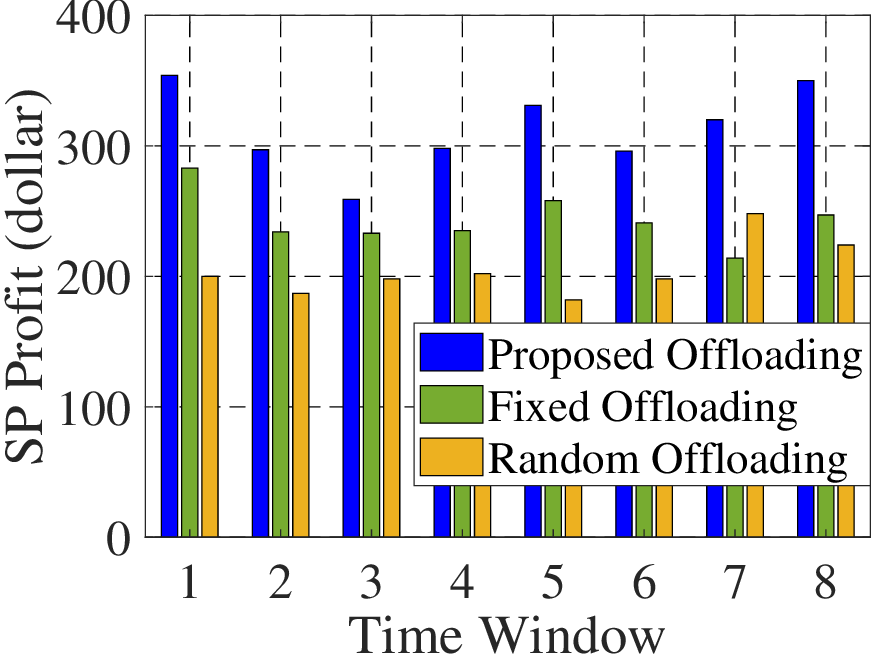}
            \vspace{-0.2in}
            \caption{Analysis of SP's profit}
            \vspace{-0.1in}
            \label{fig:second}
        \end{subfigure}
        \hfill
        \begin{subfigure}{0.3\textwidth}
            \includegraphics[width=\textwidth]{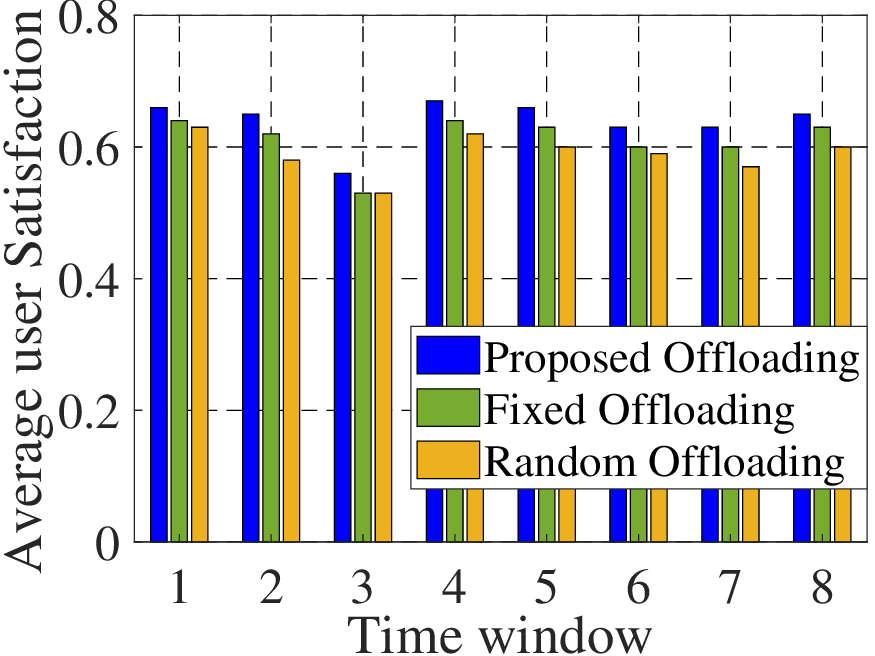}
            \vspace{-0.2in}
            \caption{Analysis of user satisfaction}
            \vspace{-0.1in}
            \label{fig:third}
        \end{subfigure}   
\caption{System Performance with Real-world Data}
\vspace*{-0.2in}
\label{systemPerformance_realworld_data}
\end{figure*}

Fig. (\ref{systemPerformance_realworld_data}) shows the evaluation results of performance metrics for different schemes with $N_U$ = 20 and $N_S$ = 40. Fig. \ref{systemPerformance_realworld_data}(a) compares the percentage of served IoTs of the proposed offloading with the benchmark schemes accross different time windows, while Fig. \ref{systemPerformance_realworld_data}(b) and \ref{systemPerformance_realworld_data}(c) compare the SP profit and average user satisfaction over these intervals, respectively. The proposed strategy outperforms both the fixed and the random offloading in terms of the percentage of served IoTs, SP profit, and average user satisfaction. It achieves an average of 10\% and 28\% higher served percentage than the fixed and random strategy, respectively. Similarly, the SP profit with the proposed strategy becomes 1.2 times and 1.5 times higher than the fixed and random strategy, respectively.    

Overall, our proposed offloading solution for a UAV-assisted MEC system provides a higher percentage of served IoTs than the benchmark schemes. Since it serves a larger number of IoT users, it is suitable for the urban IoT networks. Furthermore, it improves the SP revenue by an average of 19\%  along with an improvement in user satisfaction (by 12\%), thereby satisfying the needs of both users as well as the SP. Remarkably, the number of required UAVs
to achieve this performance improvement is nearly 25\% lesser in the proposed scheme compared to the benchmark scheme. 
\vspace{-2mm}
\section{Conclusion}
In this article, we have presented the UAV-assisted MEC architecture that supports IoT task offloading in urban environments. Our proposed architecture efficiently fulfills the latency and computational demands of IoT devices by deploying UAVs as relay nodes and utilizing ground ESs, combining both to achieve a robust solution that maximizes user satisfaction and SP profit. 
We use a three-sided matching optimization and an efficient UAV network deployment strategy to find a stable match between IoTs, UAVs, and ESs while utilizing the the minimum number of UAVs. 
Our proposed  UAV-assisted MEC based IoT task offloading scheme is effective in urban environments and offers an average improvement of 19\%, 12\%, and 25\% in the percentage of served IoTs, average user satisfaction, and SP profit, respectively, as compared to its benchmarks.\vspace{-0.1in} 
\vspace{-2mm}
\bibliographystyle{ieeetr}
\bibliography{TNSM_Final.bib}
\vspace{-1mm}
\begin{IEEEbiography}[{\includegraphics[width=1in,height=1.25in,clip,keepaspectratio]{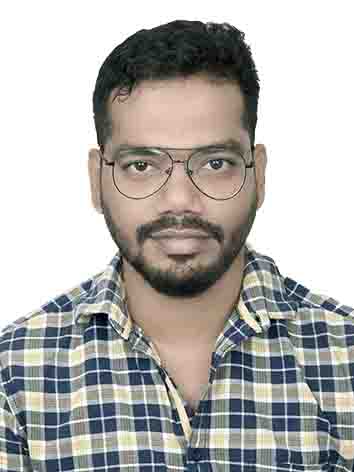}}]{Subhrajit Barick} 
received his B.Tech. degree in Electronics and Telecommunications Engineering from Biju Patnaik University of Technology (BPUT), Odisha, India in 2013, the M.Tech. degree in Communication System Engineering from Indian Institute of Information Technology, Design and Manufacturing (IIITDM), India, in 2016. He has submitted his Ph.D. thesis in the department of Electronics and Electrical Communication Engineering, at IIT Kharagpur, India. Currently he is working as an Assistant Professor in the Manipal Institute of Technology, MAHE, India. His research interests include next-generation wireless network, UAV-assisted communication, network resource management, and multi-access edge computing.
\end{IEEEbiography}
\begin{IEEEbiography}[{\includegraphics[width=1in,height=1.25in,clip,keepaspectratio]{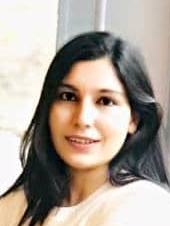}}]{Chetna Singhal [M'15, SM'21]}  is a Researcher at INRIA, France. She worked as an Assistant Professor at Indian Institute of Technology (IIT) Kharagpur from 2015 till 2024. She was an ERCIM Fellow at RI.SE, Sweden in 2022 and a visiting researcher at Northeastern University, Boston in 2018 and Politecnico di Torino, Italy in 2015 and 2016. She received her M.Tech. and Ph.D. from IIT Delhi in 2010 and 2015, respectively. Her research interests include UAV networks, machine learning orchestration, and IoT systems. She regularly serves as a TPC member and reviewer for several top-tier ACM and IEEE conferences
and journals.
\end{IEEEbiography}

\end{document}